\documentclass[aip,graphicx]{revtex4-1}
\usepackage{lineno,hyperref}
\usepackage{multirow}
\usepackage[table,xcdraw]{xcolor}
\usepackage{subcaption}
\usepackage{amsmath}
\usepackage{graphicx, tabularx}
\usepackage{float}
\usepackage{fdsymbol}

\usepackage{dcolumn}

\definecolor{sinopia}{rgb}{0.8, 0.25, 0.04}
\definecolor{steelblue}{rgb}{0.27, 0.51, 0.71}
\definecolor{brightpink}{rgb}{1.0, 0.0, 0.5}
\definecolor{royalpurple}{rgb}{0.47, 0.32, 0.66}
\definecolor{mikadoyellow}{rgb}{1.0, 0.77, 0.05}
\definecolor{princetonorange}{rgb}{1.0, 0.56, 0.0}
\definecolor{darkgreen}{rgb}{0.0, 0.25, 0.04}

\def\We{\mathrm{We}}
\def\Re{\mathrm{Re}}
\def\Oh{\mathrm{Oh}}
\def\deg{$^\circ$C}

\draft 

\begin{document}


\title{Spreading of a droplet impacting on a smooth flat surface: how liquid viscosity influences the maximum spreading time and spreading ratio} 



\author{Yunus Tansu Aksoy}\email{yunus.aksoy@kuleuven.be}
\affiliation{KU Leuven, Department of Mechanical Engineering, Division of Applied Mechanics and Energy Conversion (TME), B-3001 Leuven, Belgium}
\author{Pinar Eneren}
\affiliation{KU Leuven, Department of Mechanical Engineering, Division of Applied Mechanics and Energy Conversion (TME), B-3001 Leuven, Belgium}
\author{Erin Koos}
\affiliation{KU Leuven, Department of Chemical Engineering, Soft Matter, Rheology and Technology (SMaRT), B-3001 Leuven, Belgium}
\author{Maria Rosaria Vetrano}
\affiliation{KU Leuven, Department of Mechanical Engineering, Division of Applied Mechanics and Energy Conversion (TME), B-3001 Leuven, Belgium}

\date{\today}

\begin{abstract}
Existing energy balance models, which estimate maximum droplet spreading, insufficiently capture the droplet spreading from low to high Weber and Reynolds numbers and contact angles. This is mainly due to the simplified definition of the viscous dissipation term and incomplete modeling of the maximum spreading time. In this particular research, droplet impact onto a smooth sapphire surface is studied for seven glycerol concentrations between 0\% - 100\%, and 294 data points are acquired using high-speed photography. Fluid properties such as density, surface tension, and viscosity are also measured. For the first time according to the authors' knowledge, we incorporate the fluid viscosity in the modeling of the maximum spreading time based on the recorded data. We also estimate the characteristic velocity of the viscous dissipation term in the energy balance equation. These viscosity-based characteristic scales help to formulate a more comprehensive maximum droplet spreading model. Thanks to this improvement, our model successfully fits the data available in the literature for various fluids and surfaces compared to the existing models.
\end{abstract}


\maketitle 

\section{Introduction}\label{sec:introduction}
Droplet impact onto solid surfaces is a very popular research field thanks to an extensive variety of applications with the examples including surface coating~\cite{Josserand2016}, forensic science~\cite{Laan2014}, inkjet printing~\cite{Castrejon2013, vanDam2004}, and spray cooling~\cite{Kim2007, Moreira2010, Aksoy2021, Srikar2009}. In particular, the maximum spreading ratio $\beta_{max}$, i.e., the ratio of the maximum spreading diameter $D_m$ to the initial droplet diameter $D_0$, determines the performance of these processes~\cite{Wildeman2016}. 
As the fluid and surface properties govern droplet spreading, the following non-dimensional numbers must be considered for its modeling~\cite{Mundo1995, Rioboo2001, Zhang2017, Aksoy2021, Zhang2021}: the Weber number $\We = \rho D_0 u_0^2 / \sigma$ (inertia/surface tension forces), the Reynolds number $\Re = \rho u_0 D_0 / \eta$ (inertia/viscous forces), and the Ohnesorge number $\Oh = \sqrt{\We}/\Re$ (viscous forces/inertia and surface tension forces). The fluid properties, i.e., the density, surface tension, and dynamic viscosity are represented by $\rho$, $\sigma$, and $\eta$, respectively, and the impact parameters are denoted by the droplet diameter upon impact and the impact speed, $D_0$ and $u_0$, respectively. When a droplet comes into contact with a solid surface, it can be assumed that there is a competition between the spreading and the viscous forces affecting the droplet dynamics~\cite{Biance2004}. Inertia is considered as the dominating force during the initial wetting phase, in contrast to the viscosity~\cite{Bird2008}. Moreover, although the capillarity and the inertia are the main driving forces of the early wetting dynamics, the addition of surfactants or super-spreader solutions remarkably changes the maximum spreading~\cite{Wang2013}.

Several methods, which exist in the literature for the maximum spreading ratio estimation, are either based on empirical scaling, or energy conservation. For instance, Scheller et al.~\cite{Scheller1995} introduced the correlation presented in Eq.~\ref{eq:Scheller}, that depends on both Reynolds and Ohnesorge numbers with two empirical coefficients, $A$ and $\alpha$:
\begin{equation}
    \beta_{max} = A~(\Re^2 \cdot \Oh)^{\alpha}.
    \label{eq:Scheller}
\end{equation}
Tang et al.~\cite{Tang2017} used the same scaling parameter ($\Re^2 \cdot \Oh$) and empirically found several coefficients for 5 different surface roughness values and 4 different fluids, e.g., water, decane, ethanol, and tetradecane. Sen et al.~\cite{Sen2014} applied the same methodology on the biofuel droplets to empirically model their maximum spreading on the stainless steel substrate. On the other hand, Roisman~\cite{Roisman2009} came up with a semi-empirical relation (Eq.~\ref{eq:Roisman2009}), which approximates the Navier-Stokes equations:
\begin{equation}
    \beta_{max} = 0.87~\Re^{0.2} - 0.4~\Re^{0.4}/\sqrt{\We}.
    \label{eq:Roisman2009}
\end{equation}
Other empirical models, estimating the maximum spreading ratio available in the literature, are reported in the first section of Table~\ref{tab:correlations}.

\begin{figure}[ht]
    \centering
    \includegraphics[scale=0.56]{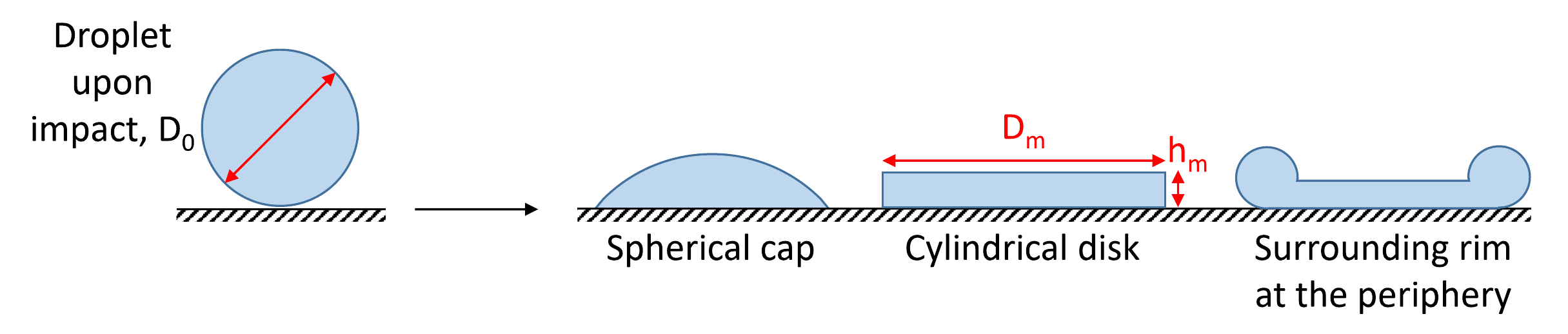}
    \caption{Geometrical approximations of the droplet shape for the surface free energy estimation at the maximum spreading.}
    \label{fig:geometries}
\end{figure}

Another strong parameter on the droplet spreading mechanism is the contact angle $\theta$ between the substrate and the droplet. Thus, models based on the energy balance better represent the physics of the phenomenon by including the contact angle as a parameter~\cite{Chandra1991}. Nevertheless, this approach requires a geometrical approximation of the droplet shape for the surface free energy estimation, as shown in Fig.~\ref{fig:geometries}. Moreover, the evaluation of viscous dissipation during the droplet spreading has to be accurately modeled. In this research, we apply the most common shape approximation, assuming the droplet as a cylinder with diameter $D_m$ and height $h_m$ at the maximum spreading, similar to these works~\cite{Pasandideh1996, Ukiwe2005, Du2021_langmuir}. In order to better incorporate the liquid-solid wettability, other geometrical approximations are also considered, such as the spherical cap model~\cite{Li2010, Park2003} or the model of the surrounding rim at the periphery of the droplets~\cite{Roisman2002, Srivastava2020, Zhang2021, Wang2019}. These energy balance models are generally tuned to best fit to some specific experimental data and may inadequately capture the droplet spreading when applied to different cases.

The energy conservation principle applied to a spreading droplet states that the sum of the kinetic energy $K_0$ and the surface free energy $S_0$ prior to the impact should be equal to the sum of the dissipated energy due to viscosity $W$, the kinetic energy $K_m$, the surface free energy $S_m$ at the maximum spreading. Since the capillary length $\lambda_c = \sqrt{(\sigma/\rho g)}$ is computed to be greater than the droplet radius, the contribution of the potential energy is neglected~\cite{Du2021, Chamakos2016}. The mathematical expression is provided in Eq.~\ref{eq:balance}:
\begin{equation}
    K_0 + S_0 = W + K_m + S_m.
    \label{eq:balance}
\end{equation}

The terms in the energy balance equation are given as follows~\cite{Pasandideh1996, Yonemoto2017, Du2021_langmuir}:
\begin{subequations}
    \begin{equation}
        K_0 = \frac{\pi \rho u_0^2 D_0^3}{12}, \label{eq:K0}
    \end{equation}
    \begin{equation}
        S_0 = \pi D_0^2 \sigma \label{eq:S0},
    \end{equation}
    \begin{equation}
        K_m = 0, \label{eq:Km}
    \end{equation}
    \begin{equation}
        S_m = \pi \sigma D_m h_m + \frac{\pi}{4} \sigma D_m^2 (1 - \cos\theta). \label{eq:Sm}
    \end{equation}
    \label{eq:terms}
\end{subequations}
The sub-equations of Eq.~\ref{eq:terms} make a perfect sphere assumption for the droplet upon impact and a stationary cylinder at the maximum spreading, which implies $h_m = 2 D_0^3 / 3 D_m^2$ due to the mass conservation~\cite{Chandra1991, Pasandideh1996}. Furthermore, different contact angle definitions appear in the literature for Eq.~\ref{eq:Sm}: equilibrium~\cite{Zhao2017}, advancing~\cite{Pasandideh1996, Ukiwe2005, Gao2014}, Young's~\cite{Chandra1991, Mao1997, Park2003, Li2010}, or contact angles at the maximum spreading~\cite{Lee2016_langmuir, Yonemoto2017, Huang2018, Wang2019}.

The $W$ term stands for the viscous dissipation in the energy balance equation, which  requires solving the following integral in Eq.~\ref{eq:W_int}~\cite{Chandra1991, Chamakos2016}:
\begin{equation}
    W = \int_{0}^{t_m} \int_{V_e}^{} \Phi dV dt,
    \label{eq:W_int}
\end{equation}
where $\Phi = \boldsymbol{\tau}: \medtriangledown \mathbf{u}$ is the viscous dissipation function~\cite{Chamakos2016}, and modeled as $\Phi \approx \eta \left(\dfrac{u_c}{L_c}\right)^2$~\cite{Chandra1991}. Therefore, the integration results in:
\begin{equation}
    W = \eta \left ( \frac{u_c}{L_c} \right )^2 V_e t_m,
    \label{eq:W}
\end{equation}
where $V_e = \dfrac{\pi D_m^2 L_c}{4}$ is the effective volume of the viscous dissipation, and $L_c$, $u_c$, and $t_m$ denote the characteristic length and velocity, and the maximum spreading time, respectively in Eq.~\ref{eq:W}. These characteristic scales represent the length and the velocity during which the viscous dissipation occurs. The maximum spreading time is defined as the time between the droplet impact and the maximum spreading, i.e., the time during which the viscous dissipation takes place. Hence, its modeling for the viscous dissipation characterization becomes particular of concern to many studies~\cite{Chandra1991, Pasandideh1996, Lee2016_langmuir, Gao2018, Qin2019}. The main models based on the energy balance approach are summarized in the second part of Table~\ref{tab:correlations} together with the chosen characteristic length, velocity, and the maximum spreading time.

Despite numerous attempts, there is still no universal equation for the prediction of the maximum droplet spreading. The goal of our study is to develop a new semi-empirical model with broader applicability. Specifically, we solve the energy balance equation by introducing new terms for the maximum spreading time and the characteristic velocity, which are functions of $\We$ and $\Re$ numbers. We perform an experimental campaign to gather a large set of data to tune our semi-empirical model. Finally, we validate the applicability of our model by using available data in the literature for various types of fluids and surfaces (large ranges of $\We$ and $\Re$ numbers, and contact angles).

\begin{table}[ht]
\caption{\label{tab:correlations} Literature models for the maximum spreading ratio $\beta_{max}$ with characteristic velocity $u_c$, characteristic length $L_c$, and maximum spreading time $t_m$.}
\scriptsize
        \begin{tabular}{lllll}
        \hline        
         Literature  & Empirical models  \\ \hline
Scheller et al.~(1995)~\cite{Scheller1995} &  $\beta_{max} = 0.61~(\Re^2 \cdot \Oh)^{0.166}$  & & &\\[5pt]

\multirow{2}{*}{Clanet et al.~(2004)~\cite{Clanet2004}}&$\beta_{max}\sim\We^{0.25}$~for~ $\We/\Re^{0.8}<1$\\
&$\beta_{max}\sim\Re^{0.2}$~for~$\We/\Re^{0.8} \ge 1$ & & &\\[5pt]

Bayer et al.~(2006)~\cite{Bayer2006} & $\beta_{max} = 0.72~(\We/\Oh)^{0.14}$ & & &\\[5pt]

Roisman et al.~(2009)~\cite{Roisman2009} & $\beta_{max} = 0.87~Re^{0.2} - 0.4~\dfrac{\Re^{0.4}}{\sqrt{\We}}$ & & &\\[5pt]

Sen et al.~(2014)~\cite{Sen2014} & $\beta_{max}=1.73(\We)^{0.14}$ & & &\\[5pt]

Laan et al.~(2014)~\cite{Laan2014} & $\beta_{max}=\dfrac{\Re^{0.2}\sqrt{P}}{A+\sqrt{P}},~P=\dfrac{\We}{\Re^{0.4}}$ & & &\\[5pt]

Seo et al.~(2015)~\cite{Seo2015} & $\beta_{max}=1.27\left(\Re^2 \cdot \Oh\right)^{0.122}$, valid for low-viscosity fluids than water. & & &\\[5pt]

Tang et al.~(2017)~\cite{Tang2017} &  $\beta_{max} = a~(\We/\Oh)^b$, a \& b depend on the fluid type and surface roughness & & &\\[6pt]
    \hline
Literature  & Energy models & $u_c$ & $L_c$ & $t_m$ \\
    \hline
    {\begin{tabular}[c]{@{}l@{}}Chandra \& \\ Avedisian~(1991)~\cite{Chandra1991} \end{tabular}} & $1.5\dfrac{\We}{\Re}\beta_{max}^4 + (1-\cos\theta)\beta_{max}^2 - \left(\dfrac{1}{3}\We+4\right)\approx 0$ & $u_0$ & $h_m$ & $\dfrac{D_0}{u_0}$\\[5pt]

    {\begin{tabular}[c]{@{}l@{}}Pasandideh-Fard \\ et al.~(1996)~\cite{Pasandideh1996} \end{tabular}} & $\beta_{max} = \sqrt{\dfrac{\We+12}{3(1-\cos\theta) + 4\dfrac{\We}{\sqrt{\Re}}}}$ & $u_0$ & $\dfrac{2D_0}{\sqrt{\Re}}$ & $\dfrac{8D_0}{3u_0}$\\[5pt]

Mao et al.~(1997)~\cite{Mao1997} & $\left[ \dfrac{1}{4} ( 1-\cos\theta ) + 0.2 \dfrac{\We^{0.83}}{\Re^{0.33}} \right] \beta_{max}^3 - \left( \dfrac{\We}{12} + 1 \right) \beta_{max} + \dfrac{2}{3} = 0$ & N/A & N/A & $\dfrac{8D_0}{3u_0}$\\[5pt]

Ukiwe et al.~(2005)~\cite{Ukiwe2005} & $(\We+12) \beta_{max} = 8 + \beta_{max}^3 \left [ 3 (1 - \cos\theta) + 4 \dfrac{\We}{\sqrt{\Re}} \right ]$ & $u_0$ & $\dfrac{2D_0}{\sqrt{\Re}}$ & $\dfrac{8D_0}{3u_0}$\\[5pt]

\multirow{2}{*}{Gao et al.~(2014)~\cite{Gao2014}}&$1 + \dfrac{\We}{12} = \dfrac{1}{6} \left ( \dfrac{D}{r_c} + \dfrac{D}{R_c} \right ) + 4 \theta_a \dfrac{r_c R_c}{D^2} + \left ( \dfrac{R_c}{D} - \dfrac{r_c}{D} \sin{\theta_a} \right )^2$ & $u_0$ & $\dfrac{2D_0}{\sqrt{\Re}}$ &$\dfrac{8D_0}{3u_0}$\\
&$+ \left ( \dfrac{R_c}{D} + \dfrac{r_c}{D} \sin{\theta_a} \right )^2 \left ( \dfrac{4}{3} \dfrac{\We}{\sqrt{\Re}} - \cos{\theta_a} \right )$ & & & \\[5pt]

    {\begin{tabular}[c]{@{}l@{}}Wildeman \\ et al.~(2016)~\cite{Wildeman2016} \end{tabular}} & $\dfrac{12}{\We}+\dfrac{1}{2} = \beta_{max}^2\dfrac{3(1-\cos\theta)}{\We} + \dfrac{\alpha}{\sqrt{\Re}}\beta_{max}^2\sqrt{\beta_{max}-1}$ no-slip, $\We>30$ & $u_0$ & $\sqrt{\dfrac{\Tilde{\tau}_m}{\Re}}$ & $\dfrac{D_0}{u_0} (\beta_{max}-1)$ \\[7pt]

\multirow{3}{*}{Huang et al.~(2018)~\cite{Huang2018}}&$\dfrac{3}{4} \left( \dfrac{\We}{\sqrt{\Re}} + \dfrac{\We^*}{\sqrt{\Re^*}} \right) \beta_{max}^4 + 3 (1 - \cos\theta) \beta_{max}^3 - (\We + 12) \beta_{max} + 8 = 0$,  for $V_0 < V^*$\\
&$\dfrac{3}{4} \left( \dfrac{\We}{\sqrt{\Re}} + \dfrac{\We^*}{\sqrt{\Re^*}} \dfrac{\Re^*}{\Re} \right) \beta_{max}^4 + 3 (1 - \cos\theta) \beta_{max}^3 - (\We + 12) \beta_{max} + 8 = 0$, for $V_0 \ge V^*$ & $u_0$ & $\dfrac{2D_0}{\sqrt{\Re}}$ & $\dfrac{D_m}{2u_0}$ \\ 
& where the superscript $^*$ denotes the critical quantity.\\[5pt]

Du et al.~(2021)~\cite{Du2021_langmuir} & $(\We+12)\beta_{max} = 8 + 3 (1-\cos\theta)\beta_{max}^3 + \left(0.98 \dfrac{\We^{1.06}}{\Re}\right)\beta_{max}^{6.5} $ & $\dfrac{3}{8}u_0$ & $\dfrac{h_m}{3}$ & $1.47 \tau_i' We^{-0.44}$\\[5pt]
    \hline
    \end{tabular}
\end{table}

\section{Experimental setup and conditions}\label{sec:experiment}
\begin{figure}[ht]
    \centering
    \includegraphics[scale=0.5]{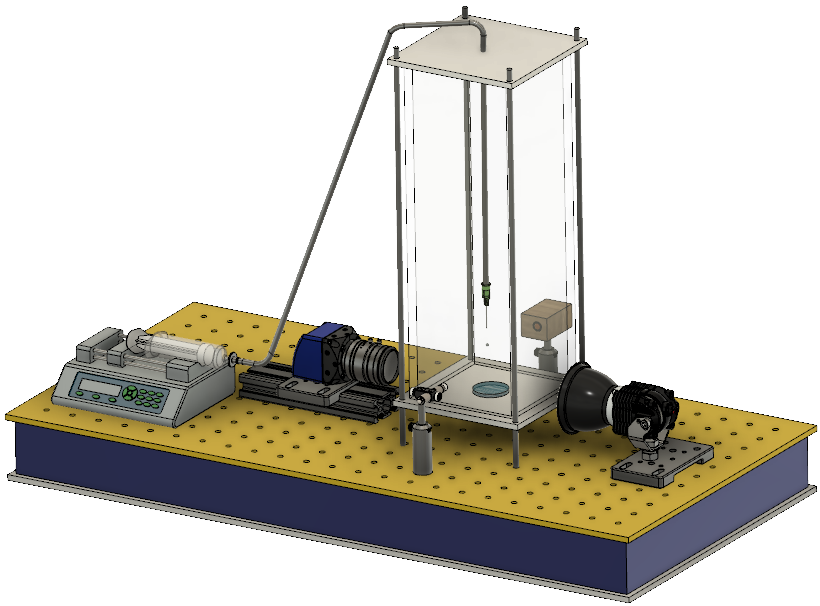}
    \caption{Illustration of the experimental setup. Syringe pump [1], fluidic tube [2], blunt-tip needle [3], sapphire substrate [4], laser-diode [5], photo-diode [6], high-speed camera [7], and LED illumination [8].}
    \label{fig:setup}
\end{figure}

The applied methodology and the experimental setup (see Fig.~\ref{fig:setup}) are described in detail in our previous work~\cite{Aksoy2022}.
Samples of various glycerol mass fractions $\omega_G$ ranging between $0 - 1$, i.e., $0~\mathrm{wt}\% - 100~\mathrm{wt}\%$ are tested. The pre-impact parameters, such as droplet diameter ($2.2~\mathrm{mm} < D_0 < 2.8~\mathrm{mm}$) and impact velocity ($1.3~\mathrm{m/s} < u_0 < 3.1~\mathrm{m/s}$), are controlled by the needle size and its position, respectively, and 294 data points are acquired ($57<\We<460$ and $4<\Re<9200$). The droplets are released on a smooth sapphire substrate with advancing contact angle of $\theta=100^\circ$. A detailed overview of the samples is given in Table~\ref{tab:properties}.
%
%

\begin{table}[ht]
\caption{\label{tab:properties} Material properties.}
\begin{tabular}{lccccc} \toprule
 \multirow{2}{*}{Sample name (label)} & $\omega_{G}$ & $\sigma$ & $\rho$ & $\eta$ \\ \cline{2-5}
& [-] & [mN/m] & [kg/m$^3$] & [mPa$\cdot$s] \\ \hline
Water (W)                & 0     & 71.3 &  996.8 &  1.0 \\
Aqueous Glycerol (AG-30) & 0.30  & 69.9 & 1071.4 &  2.5 \\
Aqueous Glycerol (AG-37) & 0.37  & 68.6 & 1090.2 &  3.3 \\
Aqueous Glycerol (AG-43) & 0.43  & 68.3 & 1105.3 &  4.1 \\
Aqueous Glycerol (AG-58) & 0.58  & 66.9 & 1143.9 &  8.7 \\
Aqueous Glycerol (AG-70) & 0.70  & 66.5 & 1177.4 & 19.7 \\
Glycerol (GLY)           & 0.999 & 62.9 & 1260.3 & 1021.5 \\ \hline
\end{tabular}
\end{table}

The droplet diameter and its impact speed are computed from the high-speed camera images by an in-house MATLAB code having a simplified particle tracking velocimetry algorithm. This code is also used for the determination of the maximum spreading ratio with an uncertainty of $\approx1.4\%$. First, the background image is subtracted from the raw images containing the droplets to have a better contrast and to remove the background objects, such as the substrate. Then, those images are converted into binary form. After the edges of the droplets are determined, pre-impact and spreading parameters are computed by using the magnification factor. The image recording properties are summarized in Table~\ref{tab:camera}.

Additionally, a circularity analysis on the droplets just before the impact yields a circularity value of $C=0.93$ based on another MATLAB routine to neglect the shape effects ($C = 4 \pi A / p^2$ where A and p are the area and perimeter of the droplet image, respectively).
\begin{table}[ht]
    \centering
    \caption{Image recording properties.}
    \label{tab:camera}
    \begin{tabular}{ccccc} \hline
        \begin{tabular}[c]{@{}c@{}}Frame rate\\ $[\mathrm{fps}]$\end{tabular} &
        \begin{tabular}[c]{@{}c@{}}Frame size\\ $[\mathrm{px}]$\end{tabular} &
        \begin{tabular}[c]{@{}c@{}}Shutter time\\ $[\mu \mathrm{s}]$\end{tabular} &
        \begin{tabular}[c]{@{}c@{}}Resolution\\ $[\mathrm{px/mm}]$\end{tabular} &
        \begin{tabular}[c]{@{}c@{}}Gray-scale level\\ $[\mathrm{bit}]$\end{tabular} \\ \hline
        12000 & 896$\times$288 & 1/35000 & 60 & 10 \\ \hline
    \end{tabular}
\end{table}
For the experimental consistency, every impact condition is repeated more than 3 times and the environmental conditions are kept constant at the temperature of $21 \pm 1$\deg~and at the atmospheric pressure. The same substrate is used to disregard the influence of the surface roughness and it is always cleaned after each droplet to prevent the presence of the dust and the remainder of the previous droplet. For that purpose, the substrate is wiped with DI water or 70\% ethanol-water mixture when necessary using an optical tissue while wearing gloves until no dust is visible in the live camera images. Moreover, it is ensured that no ethanol remains on the substrate since it affects the droplet spreading characteristics. In order to clean the needle tip and to avoid the glycerol concentration change due to air humidity, some additional droplets are discharged prior to each measurement~\cite{Palacios2013, Aksoy2022}.
\section{Evaluation of the literature models}
\begin{figure}[ht]
    \centering
    \begin{subfigure}[b]{0.49\textwidth}
        \centering
        \includegraphics[scale=0.55]{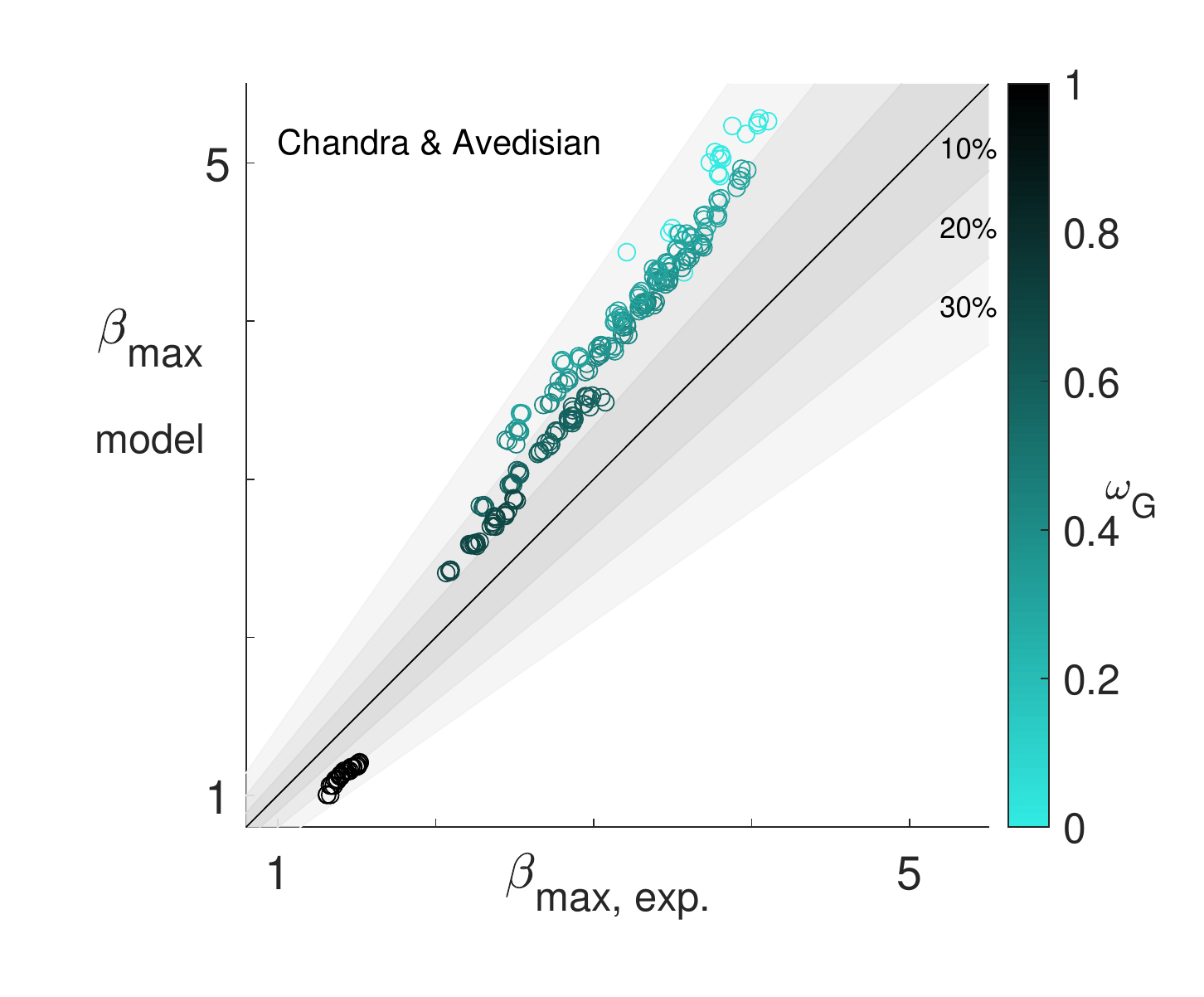}
        \caption{}
        \label{fig:chandra}
    \end{subfigure}    
    \begin{subfigure}[b]{0.49\textwidth}
        \centering
        \includegraphics[scale=0.55]{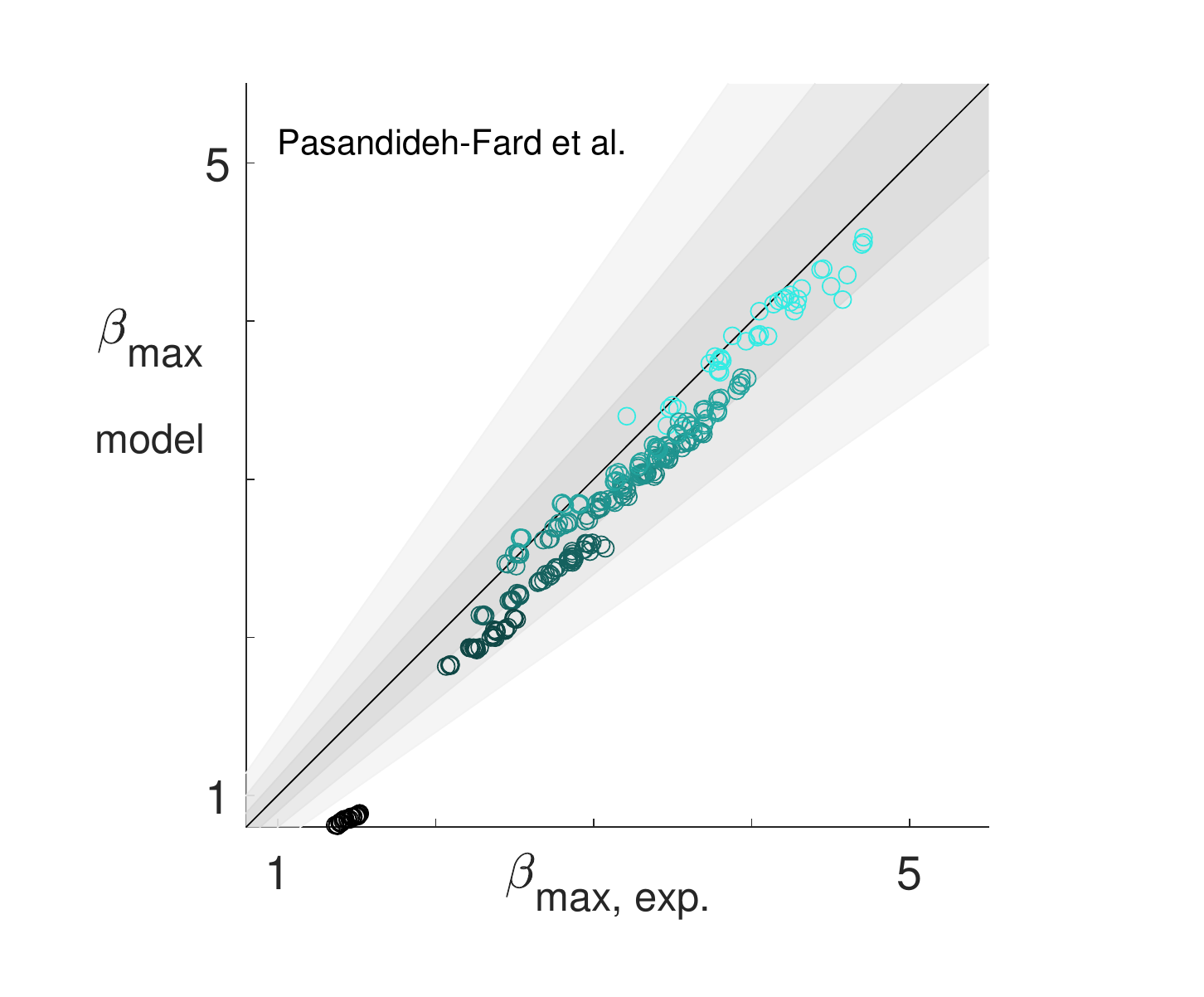}
        \caption{}
        \label{fig:pasandideh}
    \end{subfigure}
    \begin{subfigure}[b]{0.49\textwidth}
        \centering
        \includegraphics[scale=0.55]{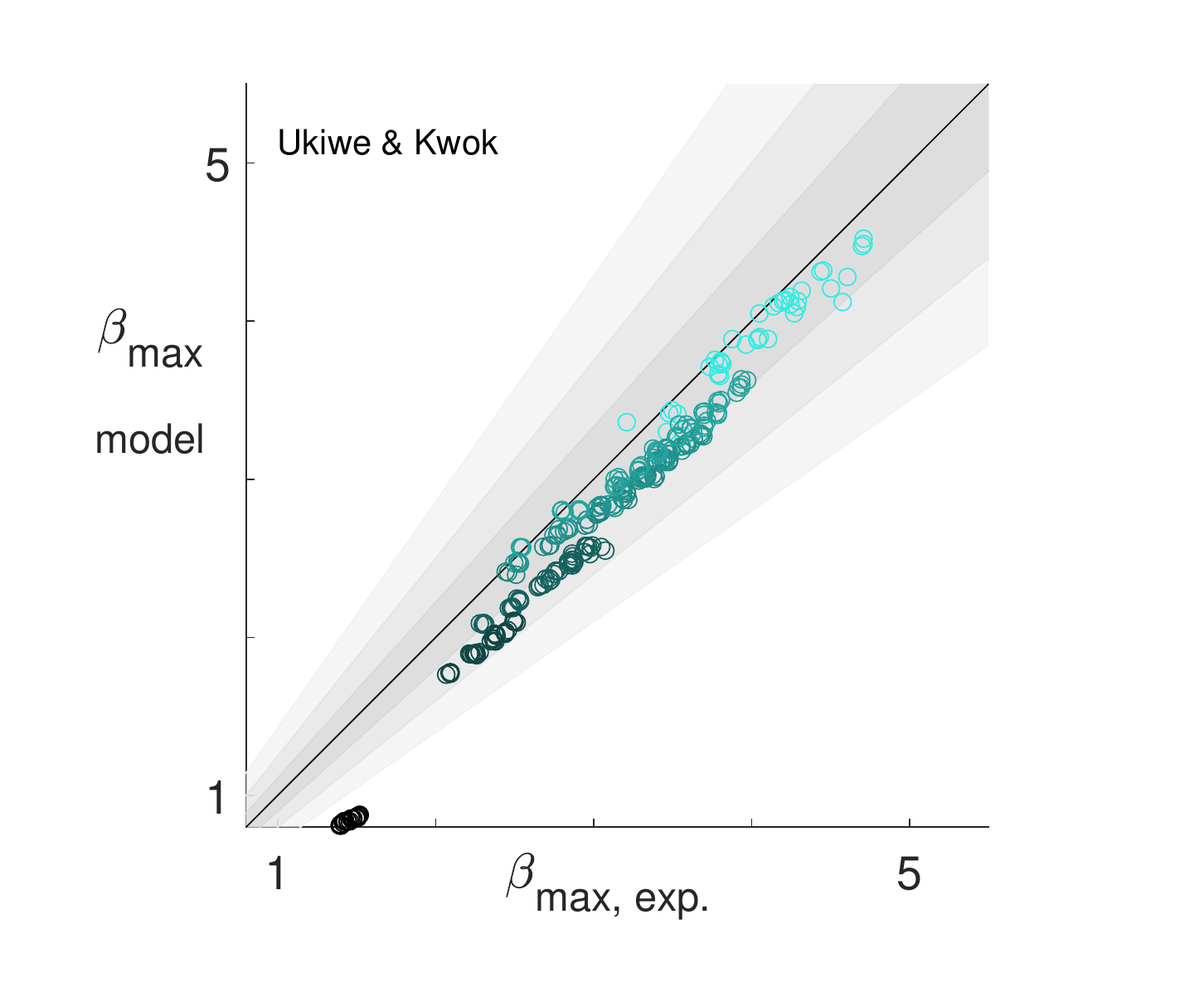}
        \caption{}
        \label{fig:ukiwe}
    \end{subfigure}
    \begin{subfigure}[b]{0.49\textwidth}
        \centering
        \includegraphics[scale=0.55]{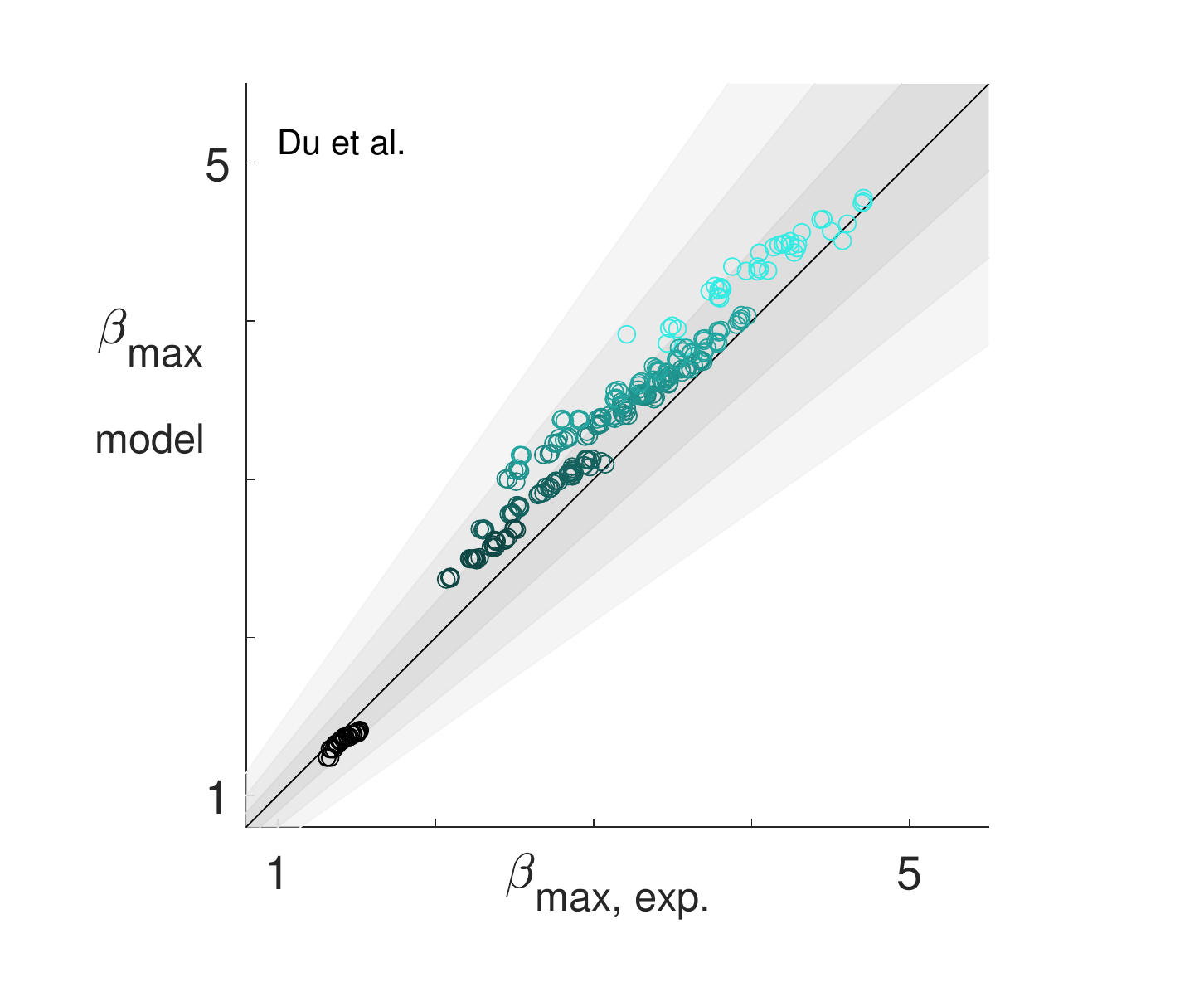}
        \caption{}
        \label{fig:du}
    \end{subfigure}    
    \caption{Performance of the several existing models on our experimental data: (a) Chandra \& Avedisian~\cite{Chandra1991}, (b) Pasandideh-Fard et al.~\cite{Pasandideh1996}, (c) Ukiwe \& Kwok~\cite{Ukiwe2005}, and (d) Du et al.~\cite{Du2021_langmuir}. The color bar is defined based on the glycerol concentration of the samples (see Table~\ref{tab:properties}). The error bands (10-30~\%, in gray color) help to present the accuracy of the models.}
    \label{fig:comparison}
\end{figure}

Fig.~\ref{fig:comparison} presents how some of the existing $\beta_{max}$ models based on the energy balance approach perform on our experimental data. Since the empirical models become ineffective for the hydrophobic and hydrophilic surfaces due to the lack of contact angle dependence, miscellaneous models based on the energy balance approach are proposed to have a better physical insight on the droplet spreading dynamics. However, it is clear that these models are only valid in the limited viscosity regimes, from which they are derived, i.e., they are not generally applicable to all the viscosity regimes. That is, if the models predict the maximum spreading ratio well for the low-viscosity liquids, they tend to underestimate it for the high-viscosity fluids, or vice versa. A very early example of the $\beta_{max}$ models based on the energy balance~\cite{Chandra1991} cannot perform well as shown in Fig.~\ref{fig:chandra}, which could be the result of selecting the improper characteristic scales. Pasandideh-Fard et al.~\cite{Pasandideh1996} changed the characteristic scales and as a consequence of that, Fig.~\ref{fig:pasandideh} represents a better agreement between the experimental results and the model for the low-viscosity liquids. By just adding the peripheral area of the cylinder to the surface free energy term, Ukiwe \& Kwok~\cite{Ukiwe2005} slightly modified the model (Fig.~\ref{fig:ukiwe}). Yet, since the viscous dissipation term in those correlations are adjusted only for the low-viscosity fluids, they still underestimate the maximum droplet spreading of the high-viscosity ones. Hence, Du et al.~\cite{Du2021_langmuir} proposed different characteristic scales and tuned the maximum spreading time based only on the high-viscosity fluids, such as aqueous glycerol and silicon oil (viscosity range of 35.5~-~220~mPa$\cdot$s). Hence, their model better perform only for the high-viscosity fluid data. On the other hand, they modeled the maximum spreading time as a function of $\We$ number only. We show in the next section that the fluid viscosity also impacts the maximum spreading time, which should be defined as a function of both $\We$ and $\Re$ numbers. The performance details of each model is discussed in Table~\ref{tab:comparison}.
\section{Model development}
The previous comparison illustrates that the models available in the literature do not sufficiently include the impact of the fluid viscosity on the determination of the maximum spreading ratio. In our model, we start from the definition of the viscous dissipation term (Eq.~\ref{eq:W}) and introduce new characteristic length and velocity, and the maximum spreading time, which are functions of both $\We$ and $\Re$ numbers to better account for the viscous nature of the fluid.
\subsection{The maximum spreading time \texorpdfstring{$t_m$}{*}}
The maximum spreading time has been modeled in several ways in the literature. Chandra and Avedisian~\cite{Chandra1991} modeled $t_m$ as the period of time for the droplet to reach zero height from its maximum value $D_0$ at a constant velocity $u_0$, i.e., $t_m = D_0 / u_0$. Pasandideh-Fard et al.~\cite{Pasandideh1996} estimated it with an additional coefficient of $8/3$ to satisfy the conservation of mass assuming the average droplet height to be $D_0/2$. Later, Lee et al.~\cite{Lee2016_langmuir} claimed that $t_m$ better scales with the maximum spreading diameter $D_m$ and propose $t_m = \frac{\sigma}{\sigma_\mathrm{ref}}\frac{D_m}{u_0}$ where $\sigma_\mathrm{ref}$ is the surface tension of the reference fluid, which is water in that case, while Huang et al.~\cite{Huang2018} used $t_m = D_m/2u_0$ to better fit their experimental data. Wildeman et al.~\cite{Wildeman2016} suggested another model including both the maximum spreading and initial droplet diameter: $t_m = (\beta_{max}-1)D_0/2u_0$. More recently, Lin et al.~\cite{Lin2018} presented an empirical correlation (Eq.~\ref{eq:Lin2018}) for the dimensionless time that fits their experimental data over wide ranges of $\We$ numbers and contact angles:
\begin{equation}
    t_m / \tau_i' = a \We^{b},
    \label{eq:Lin2018}
\end{equation}
where $a = 0.92$ and $b = -0.43$ are empirically determined constants and the modified capillary-inertial time is defined as $\tau_i'=\sqrt{\rho D_m^3 / 8 \sigma}$~\cite{Biance2004, Wang2015}. Nonetheless, in a recent publication, Du et al.~\cite{Du2021_langmuir} have computed an error of 30\% with those constants in the viscous regime. Hence, they follow the same linear regression analysis to find new constants, which are better fitting their experimental data, and they end up with slightly different values ($a = 1.47$ and $b = -0.44$).

\begin{figure}[H]
    \centering
    \begin{subfigure}[b]{0.49\textwidth}
        \centering
        \includegraphics[scale=0.55]{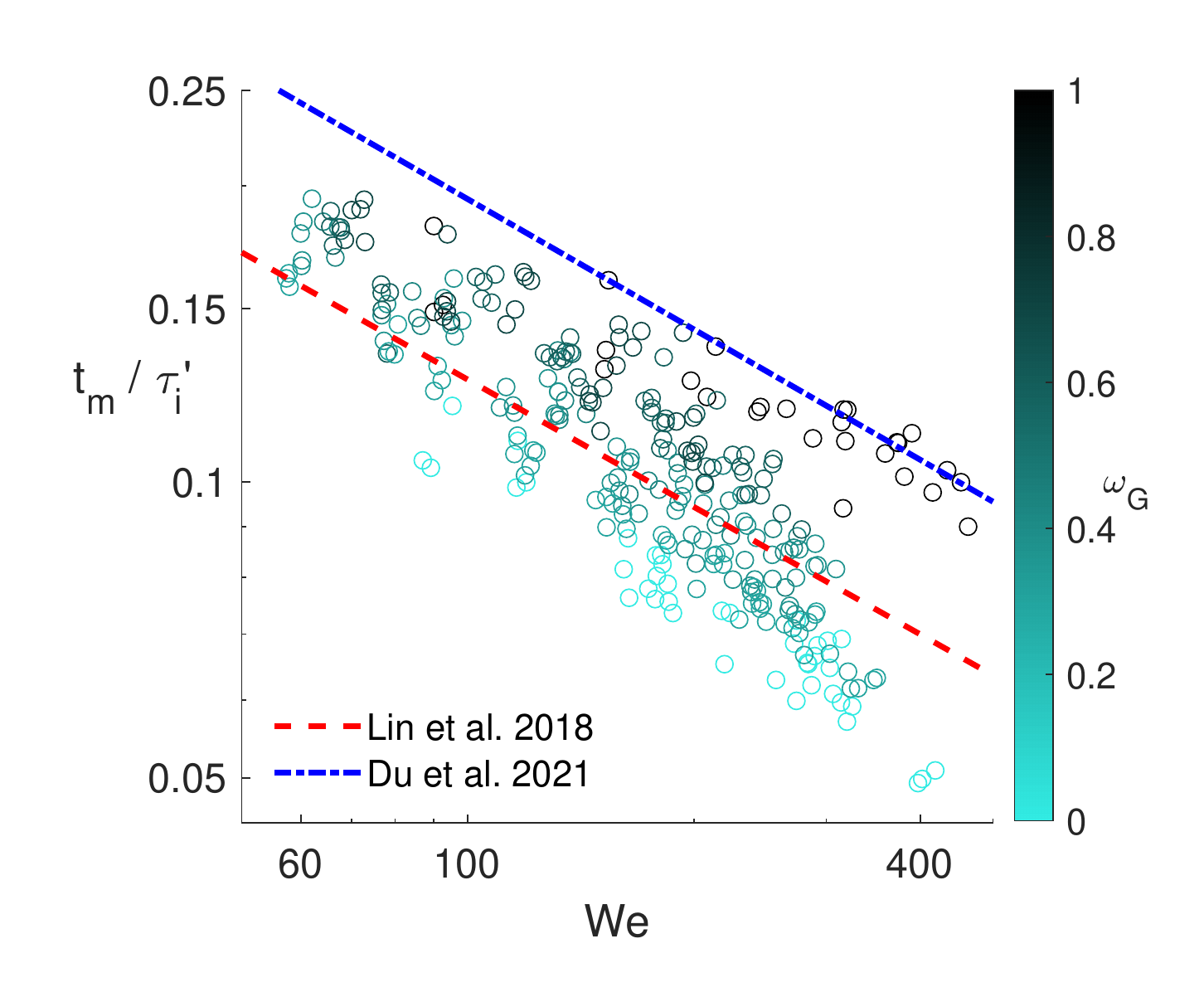}
        \caption{}
        \label{fig:time_literature}
    \end{subfigure}
   \begin{subfigure}[b]{0.49\textwidth}
        \centering
        \includegraphics[scale=0.55]{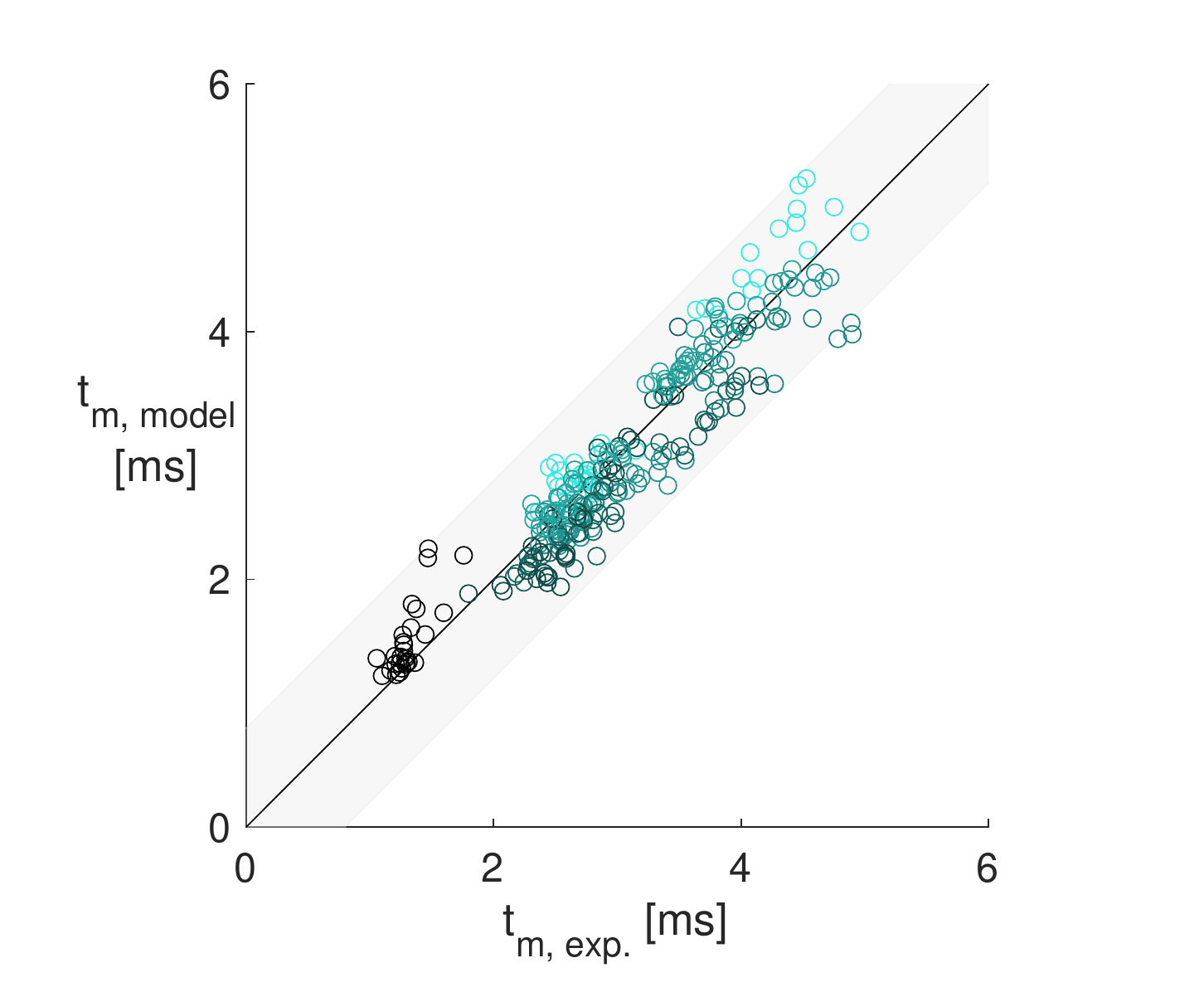}
        \caption{}
        \label{fig:time_exp-model}
    \end{subfigure}
    \caption{(a) The measured normalized maximum spreading time data with Eq.~\ref{eq:Lin2018} with constants from~\cite{Lin2018} (red dashed line) and from~\cite{Du2021_langmuir} (blue dash-dotted line). (b) Comparison of the measured and modeled (Eq.~\ref{eq:tm}) maximum spreading time. The uncertainties (0.8~ms) are represented with the light-gray band.}
    \label{fig:ts}
\end{figure}

In Fig.~\ref{fig:time_literature}, the two models of the normalized maximum spreading time $t_m/\tau_i'$ are plotted with the current experimental data as a function of $\We$ number. As our data points are colored based on the glycerol mass fraction, the viscosity dependence of the maximum spreading time becomes obvious. In other words, the scattering of the data is not random, but follows a trend: low-viscosity fluids stay below the curve of Lin et al.~\cite{Lin2018}, whereas high-viscosity samples locate at the upper side of that curve. In addition, the curve of Du et al.~\cite{Du2021_langmuir} fits the trend for the higher viscosity fluids, which more closely matches their data set. Thus, to correctly address the influence of the viscous forces, we propose to include a dependence also on $\Re$ number:
\begin{equation}
    t_m / \tau_i'= a \We^b \Re^c,
    \label{eq:tm}
\end{equation}
using the empirically determined constants $a=2$, $b=-0.45$, and $c=-0.09$ in Eq.~\ref{eq:tm}. Consequently, the maximum spreading time model based on Eq.~\ref{eq:tm} completely fits our experimental data within the uncertainty limits, as shown in Fig.~\ref{fig:time_exp-model}.
\subsection{The characteristic length \texorpdfstring{$L_c$}{*} \& characteristic velocity \texorpdfstring{$u_c$}{*}}
Previous works have selected different characteristic length and characteristic velocity based on various postulations, which are also summarized in Table~\ref{tab:correlations}.
Several length scales have been considered to estimate the characteristic length $L_c$. For instance, Chandra and Avedisian~\cite{Chandra1991} assumed the characteristic length as the rim thickness $h_m$. On the other hand, Pasandideh-Fard et al.~\cite{Pasandideh1996} postulated that length to be equal to the boundary layer thickness, in which the viscous dissipation occurs since the previous model overestimates $\beta_{max}$ by 40\%. Yonemoto et al.~\cite{Yonemoto2017} proposed taking the harmonic average of the possible droplet shapes at the end of the spreading process (see Fig.~\ref{fig:geometries}), thus ending up with $L_c=h_m/3$. Ruiter et al.~\cite{Ruiter2010} stated that the expected boundary layer thickness always stays smaller than the rim thickness. Further, the rim thickness increases with viscosity, but decreases with increasing impact speed. Mao et al.~\cite{Mao1997} defined low and high viscosity regimes separately, and claim that these regimes cannot be characterized only with $\Re$ number, i.e., viscosity. In our work, we consider $L_c = h_m$ as it implicitly includes material and impact properties at the maximum spreading~\cite{Ruiter2010}.

\begin{figure}[ht]
    \centering
    \includegraphics[width=0.5\textwidth]{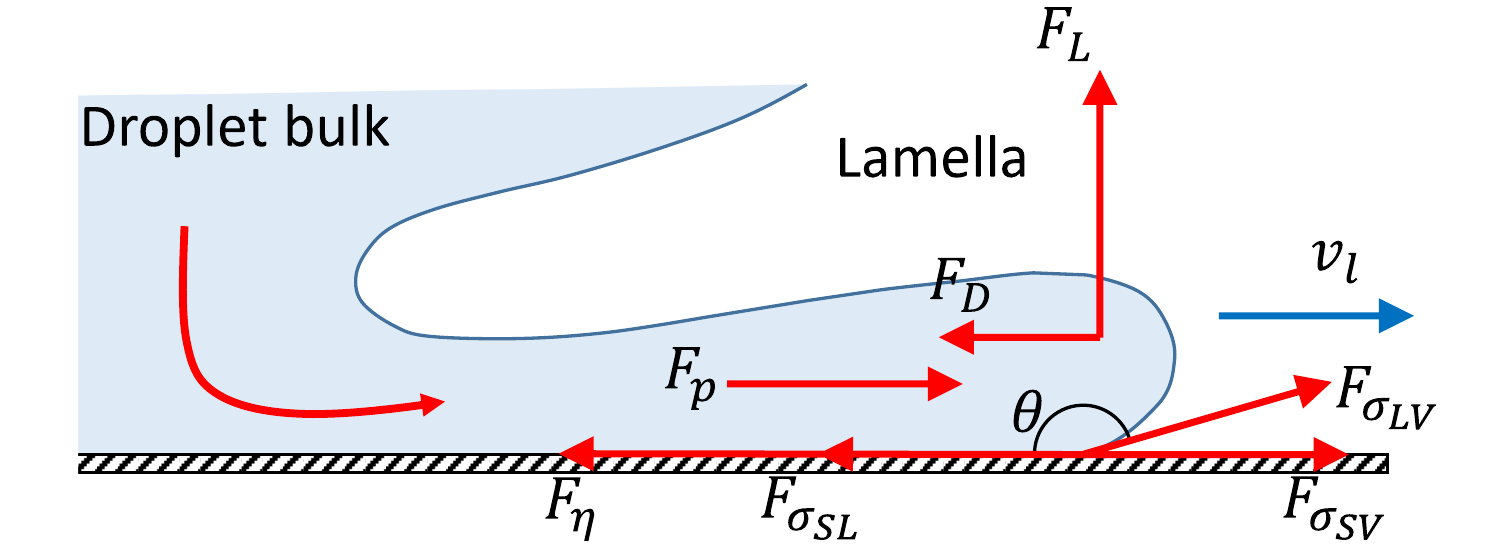}
    \caption{Forces affecting the lamella spreading speed $v_l$: propelling forces due to impact $F_p$, viscous forces $F_\eta$, forces due to surface tension $F_{\sigma_{LV}}$, lift $F_L$, and drag $F_D$.}
    \label{fig:forces}
\end{figure}

In most of the previous works applying the energy balance approach, the impact speed is used in Eq.~\ref{eq:W} as the characteristic velocity, i.e., $u_c = u_0$~\cite{Wang2019}. Nevertheless, Yonemoto et al.~\cite{Yonemoto2017} and Du et al.~\cite{Du2021_langmuir} approximated the characteristic velocity as $u_c = 3/8~u_0$ as a simplification of the radial velocity. Zhang et al.~\cite{Zhang2021} expressed that the lamella spreading speed depends on the droplet impact speed and an increased contact angle positively affects the lamella spreading speed. Although the propelling force, which is due to the droplet impact as a result of the conservation of momentum, drives the lamella spreading, the surface tension and the viscous forces must also be taken into account when the force equilibrium is considered, as illustrated in Fig.~\ref{fig:forces}~\cite{Pierce2008, Almohammadi2019}.

Therefore, the lamella characteristic velocity also depends on the viscosity and the surface tension, and can be written as a function of the $\We$ and $\Re$ numbers: 
\begin{equation}
	u_c = \mathfrak{u}(\We,\Re) \cdot u_0.
\label{eq:uc}
\end{equation}
When Eq.~\ref{eq:uc} is inserted into Eq.~\ref{eq:W}, the viscous dissipation term becomes:
\begin{equation}
    W = \frac{36\pi}{8} \frac{D_0^3 \sigma}{12 D_m} \mathfrak{u}^2 \frac{\We}{\Re} \frac{u_0}{D_0} \beta_{max}^5 t_m.
    \label{eq:W_uu2L}
\end{equation}
Then, we model $\mathfrak{u}$ in order to include the dependence on the fluid properties ($\eta$ and $\sigma$) as follows:
\begin{equation}
    \mathfrak{u} = \We^{w} \Re^{r},
    \label{eq:uc_model}
\end{equation}
where $w=-0.16$ and $r=0.12$ are determined by the linear regression analysis. The negative sign of $w$ indicates that an increase in the surface tension increases $u_c$, whereas the positive sign of $r$ implies the decelerating effect of the viscosity on $u_c$. To further elaborate, the evolution of the characteristic velocity is considered as the lamella spreading speed: the viscous forces act to impede the droplet spreading, while the surface tension forces help to promote it.
\subsection{The maximum spreading ratio \texorpdfstring{$\beta_{max}$}{*}}\label{beta_max}
\begin{figure}[ht]
    \centering
    \begin{subfigure}[b]{\textwidth}
        \centering
        \includegraphics[scale=0.5]{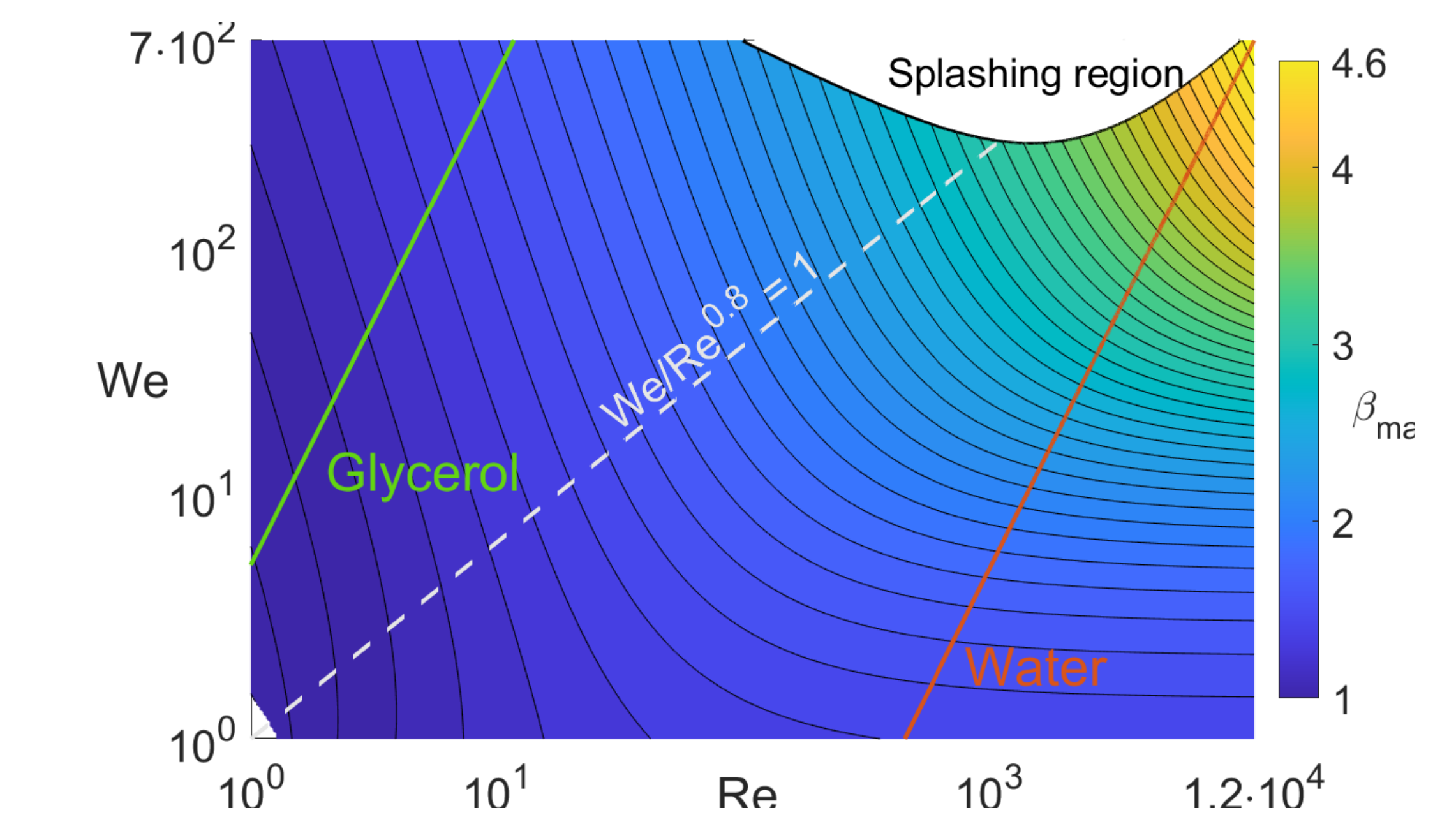}
        \caption{}
        \label{fig:WeRemap}
    \end{subfigure}\\
    \begin{subfigure}[b]{0.49\textwidth}
        \centering
        \includegraphics[scale=0.55]{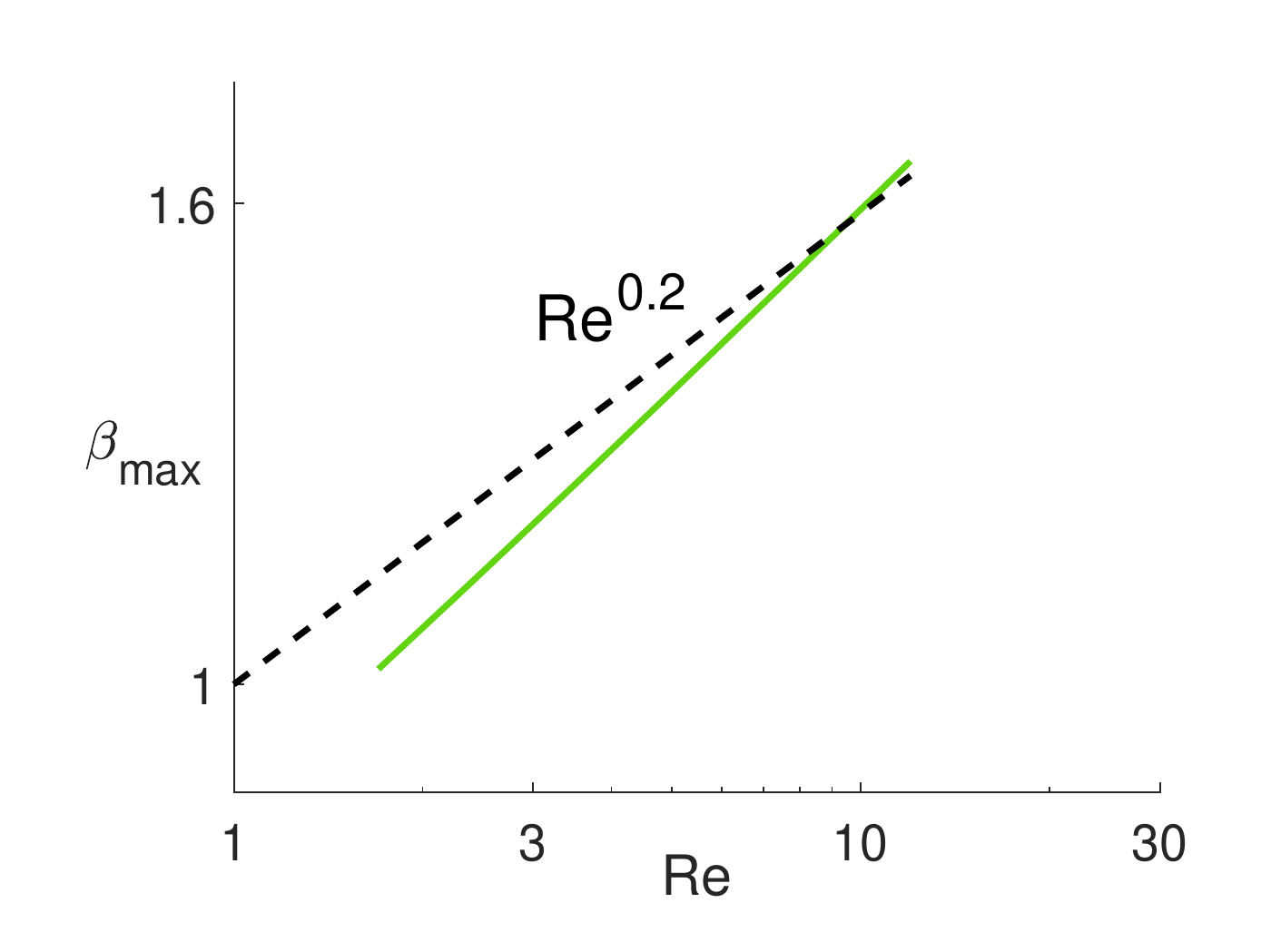}
        \caption{}
        \label{fig:ReBetamaxVis}
    \end{subfigure}
    \begin{subfigure}[b]{0.49\textwidth}
        \centering
        \includegraphics[scale=0.55]{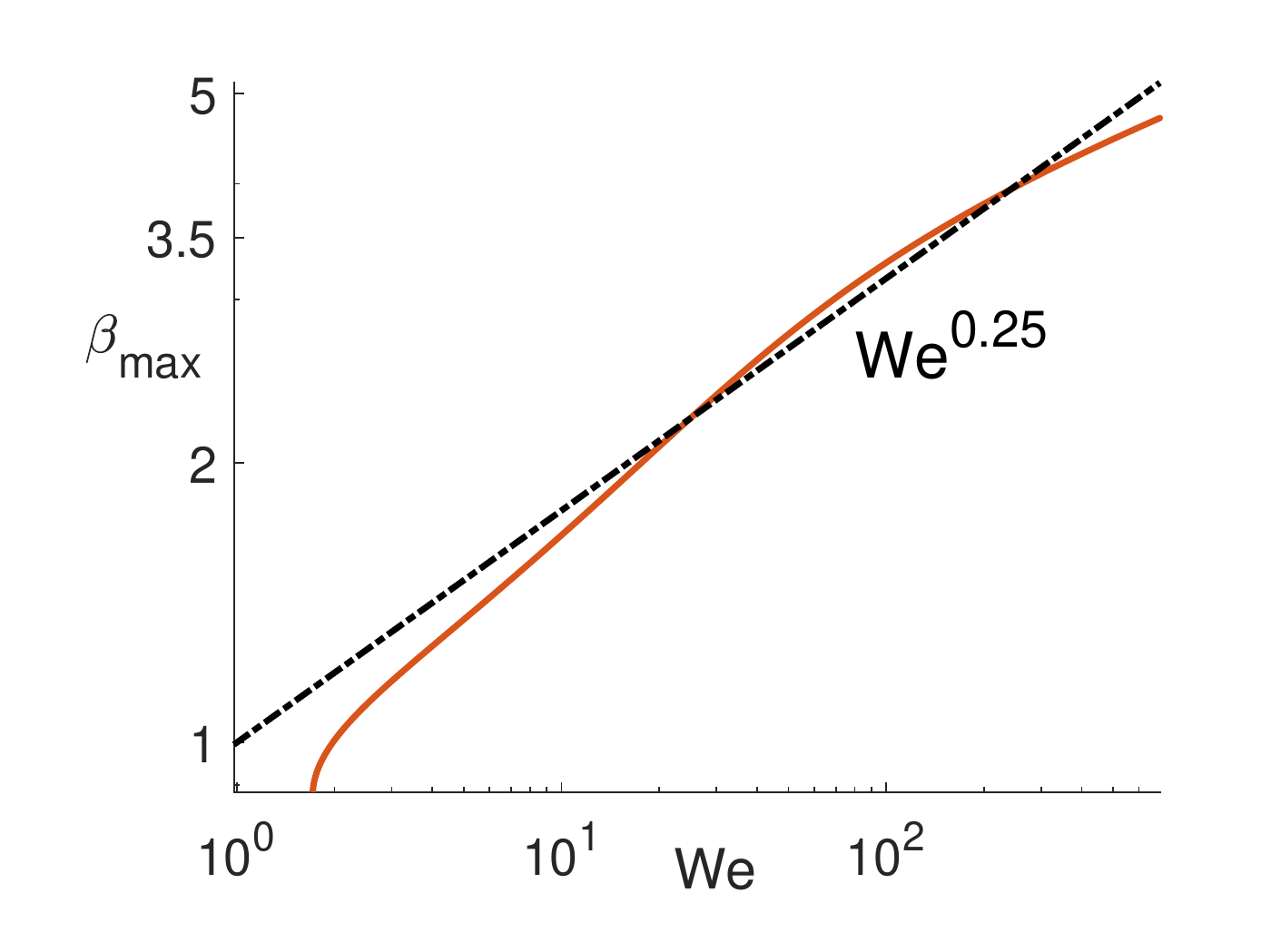}
        \caption{}
        \label{fig:WeBetamax170}
    \end{subfigure}
    \caption{Applicability of Eq.~\ref{eq:beta_final} over large ranges of $\We$ and $\Re$ numbers: (a) $\We-\Re-\beta_{max}$ contour map with $\Oh_\mathrm{W}=0.0022$ and $\Oh_\mathrm{GLY}=2.26$ (whitish gray dashed line represents the line at $\We/\Re^{0.8}=1$ and the correlation in~\cite{Aksoy2022} is applied to exclude the splashing region). Comparison of our model with the scaling laws~\cite{Clanet2004}: spreading behavior of (b) glycerol droplet on a smooth surface (dashed line for $\Re^{0.2}$) and (c) water droplet on a super-hydrophobic surface (dashed-dotted line for $\We^{0.25}$).}
    \label{fig:WeRe}
\end{figure}

\begin{figure}[ht]
    \centering
    \begin{subfigure}[b]{0.49\textwidth}
        \centering
        \includegraphics[scale=0.53]{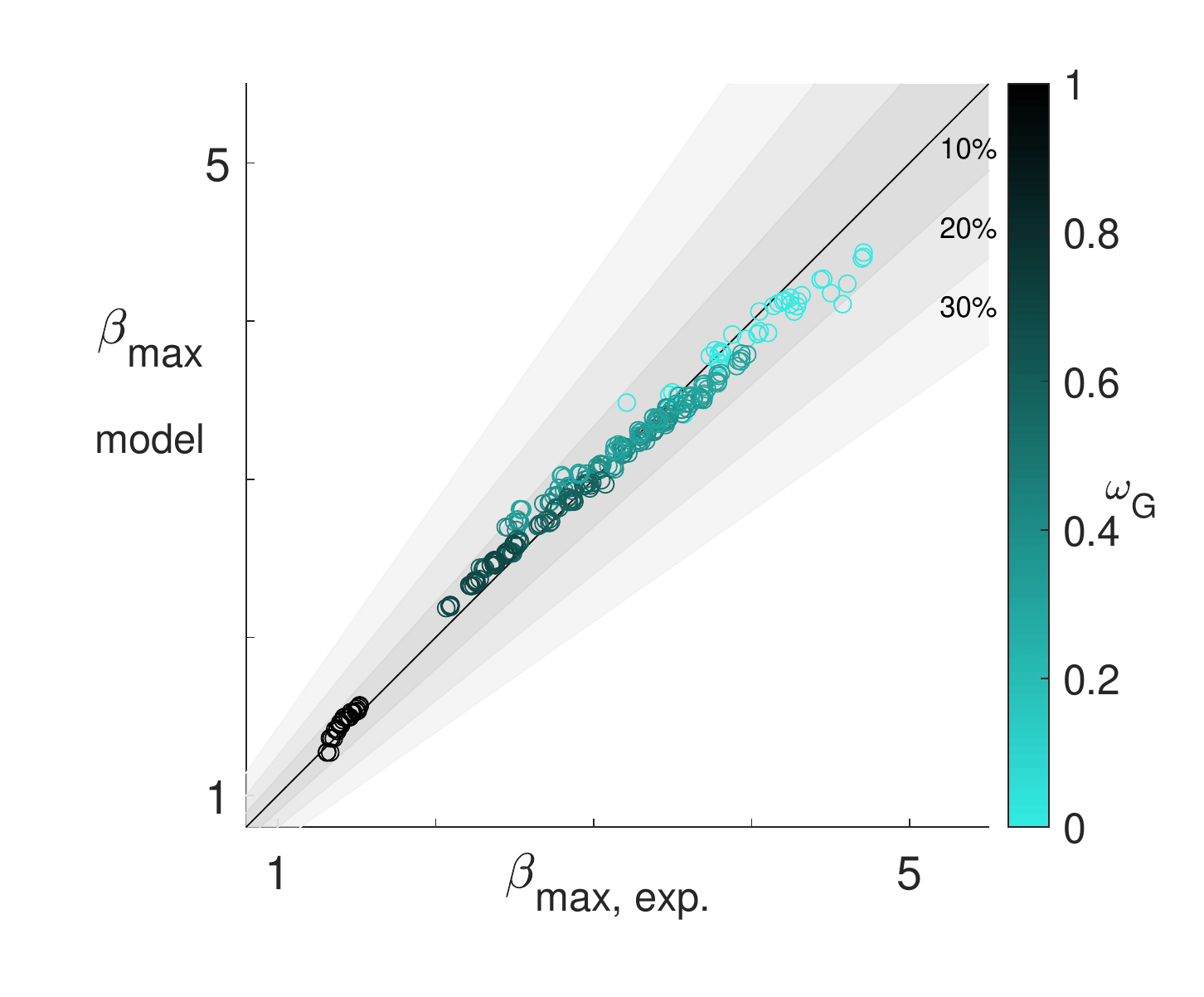}
        \caption{}
        \label{fig:new-model-data}
    \end{subfigure}
    \begin{subfigure}[b]{0.49\textwidth}
        \centering
        \includegraphics[scale=0.53]{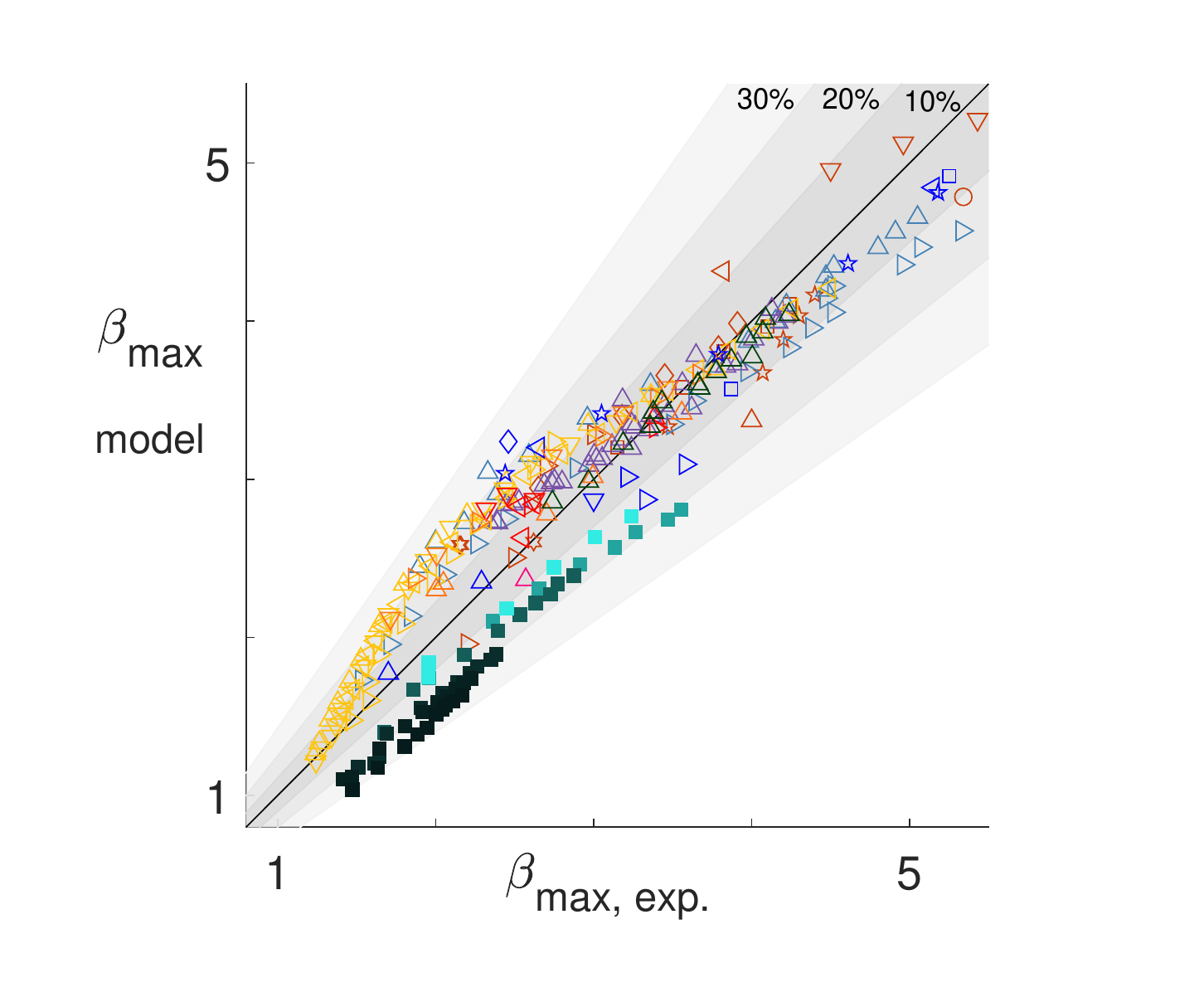}
        \caption{}
        \label{fig:new-model-literature}
    \end{subfigure}
    \caption{Assessment of the new model (Eq.~\ref{eq:beta_final}) on: (a) current experimental data (b) a very wide range of literature data. The detailed legend is given in Table~\ref{tab:legend}. Current experimental data (empty circles) and the data from~\cite{Zhao2017} (solid squares) are colored based on the glycerol concentration. The error bands (10-30~\%, in gray color) help to present the accuracy of the model.}
    \label{fig:new-model}
\end{figure}

Substituting Eq.~\ref{eq:tm} and Eq.~\ref{eq:uc_model} into Eq.~\ref{eq:W_uu2L}, we obtain the final form of the viscous dissipation term. Using Eq.~\ref{eq:balance}, the final expression of the maximum spreading becomes:
\begin{equation}
    3.18 \frac{\We^{0.72}}{\Re^{0.86}} \beta_{max}^{6.5} + 3 (1-\cos{\theta}) \beta_{max}^{3} - (\We + 12) \beta_{max} + 8 = 0.
    \label{eq:beta_final}
\end{equation}
After solving Eq.~\ref{eq:beta_final} over broad ranges of $\We$ and $\Re$ numbers, we obtain the map shown in Fig.~\ref{fig:WeRemap}. In the viscous regime, where $\We\gg\Re$, the maximum spreading ratio mainly depends on $\Re$ number and the effect of $\We$ number is negligible. Likewise, for the low-viscosity fluids, such as water with $\Re\gg\We$, there is a stronger dependence on the $\We$ number and a negligible effect of $\Re$ number. These observations, plotted in Fig.~\ref{fig:ReBetamaxVis} \&~\ref{fig:WeBetamax170}, are also in good accordance with the scaling laws proposed by Clanet et al.~\cite{Clanet2004}: $\beta_{max} \approx \Re^{0.2}$ for high-viscosity fluids ($\We/\Re^{0.8}>1$) and $\beta_{max} \approx \We^{0.25}$ for water droplet on hydrophobic surfaces.
In order to verify the wide applicability of the new model (Eq.~\ref{eq:beta_final}), we try to apply it on numerous experimental data points available in the literature. As presented in Fig.~\ref{fig:new-model}, whose legend is separately reported in Table~\ref{tab:legend}, these measurements cover large spectra of contact angles and fluid viscosities ($1<\We<1283$, $4<\Re<35000$, and $5.6^\circ<\theta<140^\circ$ \& $\theta\approx160^\circ$). More in detail, Fig.~\ref{fig:new-model-data} depicts the performance of our model on the current experimental data, from which a good agreement is anticipated. Furthermore, it is plotted on the data from literature (276 data points) in Fig.~\ref{fig:new-model-literature}, which proves its inclusiveness over quite miscellaneous cases.
\begin{table}[ht]
\centering
\caption{The details of the legend given in Fig.~\ref{fig:new-model}.}
\label{tab:legend}
\begin{tabular}{ccccc|ccccc}
\hline
Fluid & Surface & Marker & CA$^\circ$ & Ref. &
Fluid & Surface & Marker & CA$^\circ$ & Ref.\\ \hline
Heptane & Stainless steel & \textcolor{sinopia}{$\triangle$} & 20 & \cite{Pasandideh1996} &
Paraffin wax & Aluminium & \textcolor{sinopia}{$\triangleright$} & 140 & \cite{Pasandideh1996}\\
Tin & $\mathrm{Al_2O_3}$ & \textcolor{sinopia}{$\triangledown$} & 140 & \cite{Pasandideh1996} &
Tin & Stainless steel & \textcolor{sinopia}{$\triangleleft$} & 140 & \cite{Pasandideh1996} \\
Water & Beeswax & \textcolor{sinopia}{$\meddiamond$} & 111 & \cite{Pasandideh1996} &
Water & Cellulose acetate & \textcolor{sinopia}{$\square$} & 62 & \cite{Pasandideh1996} \\
Water & Glass & \textcolor{sinopia}{$\medwhitestar$} & 27 & \cite{Pasandideh1996} &
Water & Steel & \textcolor{sinopia}{$\starofdavid$} & 110 & \cite{Pasandideh1996} \\
Zinc & Stainless steel & \textcolor{sinopia}{$\medcircle$} & 140 & \cite{Pasandideh1996} &
Water & Glass & \textcolor{steelblue}{$\triangle$} & 64 & \cite{Huang2018} \\
Water & Parafilm & \textcolor{steelblue}{$\triangleright$} & 110 & \cite{Huang2018} &
Diesel & Aluminium & \textcolor{brightpink}{$\triangle$} & 83 & \cite{Jadidbonab2018} \\
Blood & Stainless steel & \textcolor{royalpurple}{$\triangle$} & 90 & \cite{Laan2014} &
Water & Meule & \textcolor{mikadoyellow}{$\triangle$} & 112 & \cite{Lee2016_JCIS}   \\
Water & Pietra serena & \textcolor{mikadoyellow}{$\triangleright$} & 99 & \cite{Lee2016_JCIS} &
Water & Savonnieres & \textcolor{mikadoyellow}{$\triangledown$} & 125 & \cite{Lee2016_JCIS} \\
Water & Steel & \textcolor{mikadoyellow}{$\triangleleft$} & 102 & \cite{Lee2016_JCIS} &
AG & Acrylic & \textcolor{red}{$\triangle$} & 127 & \cite{Santiago2020} \\
AG & Glass & \textcolor{red}{$\triangleright$} & 155 & \cite{Santiago2020} &
Water & Acrylic & \textcolor{red}{$\triangledown$} & 109 - 121& \cite{Santiago2020} \\
Water & Glass & \textcolor{red}{$\triangleleft$} & 125 - 143 & \cite{Santiago2020} &
Glycerol & Wax & \textcolor{blue}{$\triangle$} & 97 & \cite{Sikalo2002} \\
Isopropanol & Smooth glass & \textcolor{blue}{$\triangleright$} & $\approx0$ & \cite{Sikalo2002} &
Isopropanol & Wax & \textcolor{blue}{$\triangledown$} & $\approx0$ & \cite{Sikalo2002} \\
Water & PVC & \textcolor{blue}{$\triangleleft$} & 83 & \cite{Sikalo2002} &
Water & Rough glass & \textcolor{blue}{$\meddiamond$} & 78 & \cite{Sikalo2002} \\
Water & Smooth glass & \textcolor{blue}{$\square$} & 10 & \cite{Sikalo2002} &
Water & Wax & \textcolor{blue}{$\medwhitestar$} & 105 & \cite{Sikalo2002} \\
Ethanol & Silicon rubber & \textcolor{princetonorange}{$\triangle$} & 50 & \cite{Yonemoto2017} &
Water & Polycarbonate & \textcolor{princetonorange}{$\triangleright$} & 95 & \cite{Yonemoto2017} \\
Water & Silicon rubber & \textcolor{princetonorange}{$\triangledown$} & 122 & \cite{Yonemoto2017} &
Biofuel & Stainless steel & \textcolor{darkgreen}{$\triangle$} & 5.6 & \cite{Sen2014}   \\ \hline
\end{tabular}
\end{table}

Table~\ref{tab:comparison} quantifies the performances of several models from literature based on the aforementioned data. To evaluate the performance of a model, the following mean error formula is employed:
\begin{equation}
    \mathrm{mean} \left( \left|
    \frac{\beta_\mathrm{max,exp}-\beta_\mathrm{max,model}}{\beta_\mathrm{max,exp}}
    \right| \right) \times 100\%.
    \label{eq:meanerror}
\end{equation}
The experimental data both collected from the current study and extracted from the literature (see the first column of Table~\ref{tab:comparison}) are substituted in Eq.~\ref{eq:meanerror} as $\beta_\mathrm{max,exp}$. In order to determine $\beta_\mathrm{max,model}$ for the performance evaluation of each literature model and the current study, corresponding $\We$ and $\Re$ numbers, and the contact angles are employed. We split our experimental data as water, aqueous glycerol, and glycerol for a more clear demonstration of the viscosity effects. None of the previous models outperforms our model in every viscosity regimes or for different contact angles at the same time; in other words, those which seem to perform well at low-viscosity regimes underperform for the high-viscosity data, or vice versa. 

Our model proves its comprehensiveness by having the mean error being less than 15\% for all cases, except the one with very high contact angle ($\theta\approx160^\circ$). This could be due to neglecting the low surface energy to be accounted for the super-hydrophobicity~\cite{Zhao2017}. On the other hand, only 2.5\% mean error is observed for the data of Sen et al.~\cite{Sen2014}, which has biofuel as working fluid possessing $\theta = 5.6^\circ$ on the stainless steel surface.

\begin{table}[ht]
    \centering
    \caption{Performance evaluation (based on Eq.~\ref{eq:meanerror}) of all the maximum spreading models on various experimental data from Fig.~\ref{fig:new-model} ($1<\We<1283$, $4<\Re<35000$, and $5.6^\circ<\theta<140^\circ$ \& $\theta\approx160^\circ$). The first three rows show current measurements that comprise water, aqueous glycerol, and glycerol to point out the viscosity effects. The models successfully estimating $\beta_{max}$ for the low-viscosity fluids could not perform well for the high-viscosity fluids, or vice versa.}
\label{tab:comparison}
\scriptsize
\begin{tabular}{cccccccccc}
\hline
 &
  \multicolumn{9}{c}{\textbf{Models}} \\ \cline{2-10} 
  \multirow{-2}{*}{\textbf{\begin{tabular}[c]{@{}c@{}}Mean error\\ $[\%]$\end{tabular}}} &  
  \textbf{\begin{tabular}[c]{@{}c@{}}Chandra \&\\ Avedisian~\cite{Chandra1991}\end{tabular}} &
  \textbf{\begin{tabular}[c]{@{}c@{}}Scheller \&\\ Bousfield~\cite{Scheller1995}\end{tabular}} &
  \textbf{\begin{tabular}[c]{@{}c@{}}Pasandideh\\ et al.~\cite{Pasandideh1996}\\ \end{tabular}} &
  \textbf{\begin{tabular}[c]{@{}c@{}}Tang \\ et al.~\cite{Tang2017}\end{tabular}} &
  \textbf{\begin{tabular}[c]{@{}c@{}}Ukiwe \&\\ Kwok~\cite{Ukiwe2005}\end{tabular}} &
  \textbf{\begin{tabular}[c]{@{}c@{}}Roisman\\~\cite{Roisman2009}\end{tabular}} &
  \textbf{\begin{tabular}[c]{@{}c@{}}Sen et\\al.~\cite{Sen2014}\end{tabular}} &
  \textbf{\begin{tabular}[c]{@{}c@{}}Du et\\al.~\cite{Du2021_langmuir}\end{tabular}} &
  \textbf{\begin{tabular}[c]{@{}c@{}}New\\ model\end{tabular}} \\ \hline
  \textbf{\begin{tabular}[c]{@{}c@{}}Water \\ \end{tabular}} &
30.9\% & 2.3\% & 3.0\% & 4.4\% & 3.3\% & 2.1\% & 9.0\% & 7.7\% & 3.0\% \\
\textbf{\begin{tabular}[c]{@{}c@{}}Aqueous glycerol \\ \end{tabular}} &
22.8\% & 2.2\% & 8.9\% & 20.6\% & 9.7\% & 1.5\% & 19.2\% & 9.1\% & 2.9\% \\ 
\textbf{\begin{tabular}[c]{@{}c@{}}Glycerol\\  \end{tabular}} &
20.5\% & 5.8\% & 41.4\% & 51.7\% & 42.6\% & 13.3\% & 164.9\% & 5.1\% & 3.2\% \\ \hline
\textbf{\begin{tabular}[c]{@{}c@{}}Pasandideh et al.~\cite{Pasandideh1996} \end{tabular}} &
31.9\% & 11.2\% & 5.7\% & 16.1\% & 5.6\% & 9.8\% & 17.5\% & 12.1\% & 7.2\% \\ 
\textbf{\begin{tabular}[c]{@{}c@{}}Huang et al.~\cite{Huang2018} \end{tabular}} &
34.9\% & 12.5\% & 12.3\% & 6.9\% & 9.4\% & 9.7\% & 16.3\% & 16.0\% & 11.6\% \\ 
\textbf{\begin{tabular}[c]{@{}c@{}}Jadidbonab et al.~\cite{Jadidbonab2018} \end{tabular}} &
6.0\% & 14.5\% & 19.9\% & 42.1\% & 22.0\% & 13.9\% & 20.8\% & 1.2\% & 8.0\% \\ 
\textbf{\begin{tabular}[c]{@{}c@{}}Laan et al.~\cite{Laan2014} \end{tabular}} &
27.9\% & 4.2\% & 7.3\% & 12.8\% & 7.4\% & 4.9\% & 22.1\% & 11.2\% & 5.9\% \\ 
\textbf{\begin{tabular}[c]{@{}c@{}}Lee et al.~\cite{Lee2016_JCIS} \end{tabular}} &
37.9\% & 30.8\% & 19.2\% & 7.5\% & 8.3\% & 39.1\% & 30.1\% & 22.1\% & 14.7\% \\ 
\textbf{\begin{tabular}[c]{@{}c@{}}Santiago et al.~\cite{Santiago2020} \end{tabular}} &
27.8\% & 17.2\% & 6.9\% & 5.8\% & 6.0\% & 9.4\% & 14.2\% & 19.2\% & 9.3\% \\ 
\textbf{\begin{tabular}[c]{@{}c@{}}Sikalo et al.~\cite{Sikalo2002} \end{tabular}} &
29.9\% & 12.6\% & 13.0\% & 18.4\% & 12.8\% & 10.8\% & 21.9\% & 13.6\% & 10.4\% \\ 
\textbf{\begin{tabular}[c]{@{}c@{}}Yonemoto et al.~\cite{Yonemoto2017} \end{tabular}} &
38.0\% & 15.2\% & 11.8\% & 14.6\% & 8.1\% & 12.7\% & 15.1\% & 24.2\% & 11.3\% \\
\textbf{\begin{tabular}[c]{@{}c@{}}Sen et al.~\cite{Sen2014} \end{tabular}} &
33.7\% & 5.8\% & 4.8\% & 18.0\% & 4.3\% & 6.1\% & 1.3\% & 8.5\% & 2.5\% \\
\textbf{\begin{tabular}[c]{@{}c@{}}Zhao et al.~\cite{Zhao2017} \end{tabular}} &
15.5\% & 22.6\% & 32.3\% & 51.3\% & 46.0\% & 23.8\% & 35.6\% & 13.9\% & 19.1\% \\\hline
\end{tabular}
\end{table}
\section{Conclusions}\label{sec:conclusion}
After a thorough literature study, it is seen that the available maximum spreading models are only applicable to the limited fluid-surface combinations and there is a lack of general application. In particular, the models tuned for the low-viscosity fluids~\cite{Scheller1995, Tang2017, Pasandideh1996, Ukiwe2005} underestimate the maximum droplet spreading for the high-viscosity fluids, whereas others~\cite{Du2021_langmuir} overestimate it for the low-viscosity fluids while correctly predicting the high-viscosity ones. Hence, we come up with a more comprehensive model having a physical insight, which works in both high and low-viscosity regimes, and satisfyingly matches abundant literature data over broad ranges of $1<\We<1283$ and $4<\Re<35000$ numbers, as well as various surface wettability conditions ($5.6^\circ<\theta<140^\circ$ and $\theta\approx160^\circ$).

An extensive experimental campaign, composed of 294 data points in the viscosity range of 1~-~1021~mPa$ \cdot $s, is carried out. We first discover the viscosity effect in the maximum spreading time model, which improves the viscous dissipation term. Afterwards, we model the maximum spreading of the droplet following the energy balance approach by having the characteristic velocity written as a function of $\We$ and $\Re$ numbers. To validate the new model, we apply it on 276 data points from 10 previous studies~\cite{Pasandideh1996, Huang2018, Jadidbonab2018, Laan2014, Lee2016_JCIS, Santiago2020, Sikalo2002, Yonemoto2017, Zhao2017, Sen2014}. This evaluation reveals its unprecedented performance among the existing models. The mean error is less than 15\% for all the aforementioned conditions, except the super-hydrophobic case. Even though there could be better models for specific data sets, our model is the most inclusive one, proving good accordance over a wide variety of fluids and surfaces. Thanks to its high accuracy in the predictions of the maximum spreading of the droplet, the performance of the processes such as surface coating and inkjet printing, can be highly enhanced. It should be remarked that the model developed in our work uses the cylindrical shape approximation at maximum droplet spreading. Therefore, the predictions can deviate for the extreme droplet shapes, which can be overcome by improved shape approximations in future studies. In addition, for further improvements, we recommend an experimental study to directly measure the velocity profile during the droplet spreading for more appropriate modeling of the characteristic length and characteristic velocity, e.g., Particle Image Velocimetry measurements.


%
%

%

\begin{acknowledgments}
The authors gratefully acknowledge the in-kind support of Department of Chemical Engineering, KU Leuven. This work was supported by Interne Fondsen KU Leuven / Internal Funds KU Leuven (C24/18/057). The authors also acknowledge Dr. Anja Vananroye for the discussions over fluid characterizations and Prof. Pierre Colinet for the discussions over droplet spreading.
\end{acknowledgments}

\bibliography{ref}

\begin{thebibliography}{53}%
\makeatletter
\providecommand \@ifxundefined [1]{%
 \@ifx{#1\undefined}
}%
\providecommand \@ifnum [1]{%
 \ifnum #1\expandafter \@firstoftwo
 \else \expandafter \@secondoftwo
 \fi
}%
\providecommand \@ifx [1]{%
 \ifx #1\expandafter \@firstoftwo
 \else \expandafter \@secondoftwo
 \fi
}%
\providecommand \natexlab [1]{#1}%
\providecommand \enquote  [1]{``#1''}%
\providecommand \bibnamefont  [1]{#1}%
\providecommand \bibfnamefont [1]{#1}%
\providecommand \citenamefont [1]{#1}%
\providecommand \href@noop [0]{\@secondoftwo}%
\providecommand \href [0]{\begingroup \@sanitize@url \@href}%
\providecommand \@href[1]{\@@startlink{#1}\@@href}%
\providecommand \@@href[1]{\endgroup#1\@@endlink}%
\providecommand \@sanitize@url [0]{\catcode `\\12\catcode `\$12\catcode
  `\&12\catcode `\#12\catcode `\^12\catcode `\_12\catcode `\%12\relax}%
\providecommand \@@startlink[1]{}%
\providecommand \@@endlink[0]{}%
\providecommand \url  [0]{\begingroup\@sanitize@url \@url }%
\providecommand \@url [1]{\endgroup\@href {#1}{\urlprefix }}%
\providecommand \urlprefix  [0]{URL }%
\providecommand \Eprint [0]{\href }%
\providecommand \doibase [0]{http://dx.doi.org/}%
\providecommand \selectlanguage [0]{\@gobble}%
\providecommand \bibinfo  [0]{\@secondoftwo}%
\providecommand \bibfield  [0]{\@secondoftwo}%
\providecommand \translation [1]{[#1]}%
\providecommand \BibitemOpen [0]{}%
\providecommand \bibitemStop [0]{}%
\providecommand \bibitemNoStop [0]{.\EOS\space}%
\providecommand \EOS [0]{\spacefactor3000\relax}%
\providecommand \BibitemShut  [1]{\csname bibitem#1\endcsname}%
\let\auto@bib@innerbib\@empty
\bibitem [{\citenamefont {Josserand}\ and\ \citenamefont
  {Thoroddsen}(2016)}]{Josserand2016}%
  \BibitemOpen
  \bibfield  {author} {\bibinfo {author} {\bibfnamefont {C.}~\bibnamefont
  {Josserand}}\ and\ \bibinfo {author} {\bibfnamefont {S.}~\bibnamefont
  {Thoroddsen}},\ }\bibfield  {title} {\enquote {\bibinfo {title} {Drop impact
  on a solid surface},}\ }\href {\doibase 10.1146/annurev-fluid-122414-034401}
  {\bibfield  {journal} {\bibinfo  {journal} {Annual Review of Fluid
  Mechanics}\ }\textbf {\bibinfo {volume} {48}},\ \bibinfo {pages} {365--391}
  (\bibinfo {year} {2016})}\BibitemShut {NoStop}%
\bibitem [{\citenamefont {Laan}\ \emph {et~al.}(2014)\citenamefont {Laan},
  \citenamefont {de~Bruin}, \citenamefont {Bartolo}, \citenamefont
  {Josserand},\ and\ \citenamefont {Bonn}}]{Laan2014}%
  \BibitemOpen
  \bibfield  {author} {\bibinfo {author} {\bibfnamefont {N.}~\bibnamefont
  {Laan}}, \bibinfo {author} {\bibfnamefont {K.~G.}\ \bibnamefont {de~Bruin}},
  \bibinfo {author} {\bibfnamefont {D.}~\bibnamefont {Bartolo}}, \bibinfo
  {author} {\bibfnamefont {C.}~\bibnamefont {Josserand}}, \ and\ \bibinfo
  {author} {\bibfnamefont {D.}~\bibnamefont {Bonn}},\ }\bibfield  {title}
  {\enquote {\bibinfo {title} {Maximum diameter of impacting liquid
  droplets},}\ }\href {\doibase 10.1103/PhysRevApplied.2.044018} {\bibfield
  {journal} {\bibinfo  {journal} {Phys. Rev. Applied}\ }\textbf {\bibinfo
  {volume} {2}},\ \bibinfo {pages} {044018} (\bibinfo {year}
  {2014})}\BibitemShut {NoStop}%
\bibitem [{\citenamefont {Castrejon-Pita}\ \emph {et~al.}(2013)\citenamefont
  {Castrejon-Pita}, \citenamefont {Baxter}, \citenamefont {Morgan},
  \citenamefont {Temple}, \citenamefont {Martin},\ and\ \citenamefont
  {Hutchings}}]{Castrejon2013}%
  \BibitemOpen
  \bibfield  {author} {\bibinfo {author} {\bibfnamefont {J.~R.}\ \bibnamefont
  {Castrejon-Pita}}, \bibinfo {author} {\bibfnamefont {W.~R.~S.}\ \bibnamefont
  {Baxter}}, \bibinfo {author} {\bibfnamefont {J.}~\bibnamefont {Morgan}},
  \bibinfo {author} {\bibfnamefont {S.}~\bibnamefont {Temple}}, \bibinfo
  {author} {\bibfnamefont {G.~D.}\ \bibnamefont {Martin}}, \ and\ \bibinfo
  {author} {\bibfnamefont {I.~M.}\ \bibnamefont {Hutchings}},\ }\bibfield
  {title} {\enquote {\bibinfo {title} {Future, opportunities and challenges of
  inkjet technologies},}\ }\href@noop {} {\bibfield  {journal} {\bibinfo
  {journal} {Atomization and Sprays}\ }\textbf {\bibinfo {volume} {23}},\
  \bibinfo {pages} {541--565} (\bibinfo {year} {2013})}\BibitemShut {NoStop}%
\bibitem [{\citenamefont {van Dam}\ and\ \citenamefont
  {Le~Clerc}(2004)}]{vanDam2004}%
  \BibitemOpen
  \bibfield  {author} {\bibinfo {author} {\bibfnamefont {D.~B.}\ \bibnamefont
  {van Dam}}\ and\ \bibinfo {author} {\bibfnamefont {C.}~\bibnamefont
  {Le~Clerc}},\ }\bibfield  {title} {\enquote {\bibinfo {title} {Experimental
  study of the impact of an ink-jet printed droplet on a solid substrate},}\
  }\href {\doibase 10.1063/1.1773551} {\bibfield  {journal} {\bibinfo
  {journal} {Physics of Fluids}\ }\textbf {\bibinfo {volume} {16}},\ \bibinfo
  {pages} {3403--3414} (\bibinfo {year} {2004})}\BibitemShut {NoStop}%
\bibitem [{\citenamefont {Kim}(2007)}]{Kim2007}%
  \BibitemOpen
  \bibfield  {author} {\bibinfo {author} {\bibfnamefont {J.}~\bibnamefont
  {Kim}},\ }\bibfield  {title} {\enquote {\bibinfo {title} {Spray cooling heat
  transfer: The state of the art},}\ }\href {\doibase
  10.1016/j.ijheatfluidflow.2006.09.003} {\bibfield  {journal} {\bibinfo
  {journal} {International Journal of Heat and Fluid Flow}\ }\textbf {\bibinfo
  {volume} {28}},\ \bibinfo {pages} {753 -- 767} (\bibinfo {year}
  {2007})}\BibitemShut {NoStop}%
\bibitem [{\citenamefont {Moreira}, \citenamefont {Moita},\ and\ \citenamefont
  {Panao}(2010)}]{Moreira2010}%
  \BibitemOpen
  \bibfield  {author} {\bibinfo {author} {\bibfnamefont {A.}~\bibnamefont
  {Moreira}}, \bibinfo {author} {\bibfnamefont {A.}~\bibnamefont {Moita}}, \
  and\ \bibinfo {author} {\bibfnamefont {M.}~\bibnamefont {Panao}},\ }\bibfield
   {title} {\enquote {\bibinfo {title} {Advances and challenges in explaining
  fuel spray impingement: How much of single droplet impact research is
  useful?}}\ }\href {\doibase 10.1016/j.pecs.2010.01.002} {\bibfield  {journal}
  {\bibinfo  {journal} {Progress in Energy and Combustion Science}\ }\textbf
  {\bibinfo {volume} {36}},\ \bibinfo {pages} {554 -- 580} (\bibinfo {year}
  {2010})}\BibitemShut {NoStop}%
\bibitem [{\citenamefont {Aksoy}\ \emph {et~al.}(2021)\citenamefont {Aksoy},
  \citenamefont {Zhu}, \citenamefont {Eneren}, \citenamefont {Koos},\ and\
  \citenamefont {Vetrano}}]{Aksoy2021}%
  \BibitemOpen
  \bibfield  {author} {\bibinfo {author} {\bibfnamefont {Y.~T.}\ \bibnamefont
  {Aksoy}}, \bibinfo {author} {\bibfnamefont {Y.}~\bibnamefont {Zhu}}, \bibinfo
  {author} {\bibfnamefont {P.}~\bibnamefont {Eneren}}, \bibinfo {author}
  {\bibfnamefont {E.}~\bibnamefont {Koos}}, \ and\ \bibinfo {author}
  {\bibfnamefont {M.~R.}\ \bibnamefont {Vetrano}},\ }\bibfield  {title}
  {\enquote {\bibinfo {title} {The impact of nanofluids on droplet/spray
  cooling of a heated surface: A critical review},}\ }\href {\doibase
  10.3390/en14010080} {\bibfield  {journal} {\bibinfo  {journal} {Energies}\
  }\textbf {\bibinfo {volume} {14}} (\bibinfo {year} {2021}),\
  10.3390/en14010080}\BibitemShut {NoStop}%
\bibitem [{\citenamefont {Srikar}\ \emph {et~al.}(2009)\citenamefont {Srikar},
  \citenamefont {Gambaryan-Roisman}, \citenamefont {Steffes}, \citenamefont
  {Stephan}, \citenamefont {Tropea},\ and\ \citenamefont {Yarin}}]{Srikar2009}%
  \BibitemOpen
  \bibfield  {author} {\bibinfo {author} {\bibfnamefont {R.}~\bibnamefont
  {Srikar}}, \bibinfo {author} {\bibfnamefont {T.}~\bibnamefont
  {Gambaryan-Roisman}}, \bibinfo {author} {\bibfnamefont {C.}~\bibnamefont
  {Steffes}}, \bibinfo {author} {\bibfnamefont {P.}~\bibnamefont {Stephan}},
  \bibinfo {author} {\bibfnamefont {C.}~\bibnamefont {Tropea}}, \ and\ \bibinfo
  {author} {\bibfnamefont {A.}~\bibnamefont {Yarin}},\ }\bibfield  {title}
  {\enquote {\bibinfo {title} {Nanofiber coating of surfaces for
  intensification of drop or spray impact cooling},}\ }\href {\doibase
  10.1016/j.ijheatmasstransfer.2009.07.021} {\bibfield  {journal} {\bibinfo
  {journal} {International Journal of Heat and Mass Transfer}\ }\textbf
  {\bibinfo {volume} {52}},\ \bibinfo {pages} {5814--5826} (\bibinfo {year}
  {2009})}\BibitemShut {NoStop}%
\bibitem [{\citenamefont {Wildeman}\ \emph {et~al.}(2016)\citenamefont
  {Wildeman}, \citenamefont {Visser}, \citenamefont {Sun},\ and\ \citenamefont
  {Lohse}}]{Wildeman2016}%
  \BibitemOpen
  \bibfield  {author} {\bibinfo {author} {\bibfnamefont {S.}~\bibnamefont
  {Wildeman}}, \bibinfo {author} {\bibfnamefont {C.~W.}\ \bibnamefont
  {Visser}}, \bibinfo {author} {\bibfnamefont {C.}~\bibnamefont {Sun}}, \ and\
  \bibinfo {author} {\bibfnamefont {D.}~\bibnamefont {Lohse}},\ }\bibfield
  {title} {\enquote {\bibinfo {title} {On the spreading of impacting drops},}\
  }\href {\doibase 10.1017/jfm.2016.584} {\bibfield  {journal} {\bibinfo
  {journal} {Journal of Fluid Mechanics}\ }\textbf {\bibinfo {volume} {805}},\
  \bibinfo {pages} {636–655} (\bibinfo {year} {2016})}\BibitemShut {NoStop}%
\bibitem [{\citenamefont {Mundo}, \citenamefont {Sommerfeld},\ and\
  \citenamefont {Tropea}(1995)}]{Mundo1995}%
  \BibitemOpen
  \bibfield  {author} {\bibinfo {author} {\bibfnamefont {C.}~\bibnamefont
  {Mundo}}, \bibinfo {author} {\bibfnamefont {M.}~\bibnamefont {Sommerfeld}}, \
  and\ \bibinfo {author} {\bibfnamefont {C.}~\bibnamefont {Tropea}},\
  }\bibfield  {title} {\enquote {\bibinfo {title} {Droplet-wall collisions:
  Experimental studies of the deformation and breakup process},}\ }\href
  {\doibase 10.1016/0301-9322(94)00069-V} {\bibfield  {journal} {\bibinfo
  {journal} {International Journal of Multiphase Flow}\ }\textbf {\bibinfo
  {volume} {21}},\ \bibinfo {pages} {151 -- 173} (\bibinfo {year}
  {1995})}\BibitemShut {NoStop}%
\bibitem [{\citenamefont {Rioboo}, \citenamefont {Tropea},\ and\ \citenamefont
  {Marengo}(2001)}]{Rioboo2001}%
  \BibitemOpen
  \bibfield  {author} {\bibinfo {author} {\bibfnamefont {R.}~\bibnamefont
  {Rioboo}}, \bibinfo {author} {\bibfnamefont {C.}~\bibnamefont {Tropea}}, \
  and\ \bibinfo {author} {\bibfnamefont {M.}~\bibnamefont {Marengo}},\
  }\bibfield  {title} {\enquote {\bibinfo {title} {Outcomes from a drop impact
  on solid surfaces},}\ }\href {\doibase 10.1615/AtomizSpr.v11.i2.40}
  {\bibfield  {journal} {\bibinfo  {journal} {Atomization and Sprays}\ }\textbf
  {\bibinfo {volume} {11}},\ \bibinfo {pages} {155--165} (\bibinfo {year}
  {2001})}\BibitemShut {NoStop}%
\bibitem [{\citenamefont {Zhang}\ \emph {et~al.}(2017)\citenamefont {Zhang},
  \citenamefont {Li}, \citenamefont {Guo},\ and\ \citenamefont
  {Lv}}]{Zhang2017}%
  \BibitemOpen
  \bibfield  {author} {\bibinfo {author} {\bibfnamefont {B.}~\bibnamefont
  {Zhang}}, \bibinfo {author} {\bibfnamefont {J.}~\bibnamefont {Li}}, \bibinfo
  {author} {\bibfnamefont {P.}~\bibnamefont {Guo}}, \ and\ \bibinfo {author}
  {\bibfnamefont {Q.}~\bibnamefont {Lv}},\ }\bibfield  {title} {\enquote
  {\bibinfo {title} {Experimental studies on the effect of reynolds and weber
  numbers on the impact forces of low-speed droplets colliding with a solid
  surface},}\ }\href {\doibase 10.1007/s00348-017-2413-z} {\bibfield  {journal}
  {\bibinfo  {journal} {Experiments in Fluids}\ }\textbf {\bibinfo {volume}
  {58}},\ \bibinfo {pages} {125} (\bibinfo {year} {2017})}\BibitemShut
  {NoStop}%
\bibitem [{\citenamefont {Zhang}\ \emph {et~al.}(2021)\citenamefont {Zhang},
  \citenamefont {Zhang}, \citenamefont {Yi}, \citenamefont {He}, \citenamefont
  {Niu},\ and\ \citenamefont {Hao}}]{Zhang2021}%
  \BibitemOpen
  \bibfield  {author} {\bibinfo {author} {\bibfnamefont {H.}~\bibnamefont
  {Zhang}}, \bibinfo {author} {\bibfnamefont {X.}~\bibnamefont {Zhang}},
  \bibinfo {author} {\bibfnamefont {X.}~\bibnamefont {Yi}}, \bibinfo {author}
  {\bibfnamefont {F.}~\bibnamefont {He}}, \bibinfo {author} {\bibfnamefont
  {F.}~\bibnamefont {Niu}}, \ and\ \bibinfo {author} {\bibfnamefont
  {P.}~\bibnamefont {Hao}},\ }\bibfield  {title} {\enquote {\bibinfo {title}
  {Effect of wettability on droplet impact: Spreading and splashing},}\ }\href
  {\doibase 10.1016/j.expthermflusci.2021.110369} {\bibfield  {journal}
  {\bibinfo  {journal} {Experimental Thermal and Fluid Science}\ }\textbf
  {\bibinfo {volume} {124}},\ \bibinfo {pages} {110369} (\bibinfo {year}
  {2021})}\BibitemShut {NoStop}%
\bibitem [{\citenamefont {Biance}, \citenamefont {Clanet},\ and\ \citenamefont
  {Qu\'er\'e}(2004)}]{Biance2004}%
  \BibitemOpen
  \bibfield  {author} {\bibinfo {author} {\bibfnamefont {A.-L.}\ \bibnamefont
  {Biance}}, \bibinfo {author} {\bibfnamefont {C.}~\bibnamefont {Clanet}}, \
  and\ \bibinfo {author} {\bibfnamefont {D.}~\bibnamefont {Qu\'er\'e}},\
  }\bibfield  {title} {\enquote {\bibinfo {title} {First steps in the spreading
  of a liquid droplet},}\ }\href {\doibase 10.1103/PhysRevE.69.016301}
  {\bibfield  {journal} {\bibinfo  {journal} {Phys. Rev. E}\ }\textbf {\bibinfo
  {volume} {69}},\ \bibinfo {pages} {016301} (\bibinfo {year}
  {2004})}\BibitemShut {NoStop}%
\bibitem [{\citenamefont {Bird}, \citenamefont {Mandre},\ and\ \citenamefont
  {Stone}(2008)}]{Bird2008}%
  \BibitemOpen
  \bibfield  {author} {\bibinfo {author} {\bibfnamefont {J.~C.}\ \bibnamefont
  {Bird}}, \bibinfo {author} {\bibfnamefont {S.}~\bibnamefont {Mandre}}, \ and\
  \bibinfo {author} {\bibfnamefont {H.~A.}\ \bibnamefont {Stone}},\ }\bibfield
  {title} {\enquote {\bibinfo {title} {Short-time dynamics of partial
  wetting},}\ }\href {\doibase 10.1103/PhysRevLett.100.234501} {\bibfield
  {journal} {\bibinfo  {journal} {Phys. Rev. Lett.}\ }\textbf {\bibinfo
  {volume} {100}},\ \bibinfo {pages} {234501} (\bibinfo {year}
  {2008})}\BibitemShut {NoStop}%
\bibitem [{\citenamefont {Wang}\ \emph {et~al.}(2013)\citenamefont {Wang},
  \citenamefont {Chen}, \citenamefont {Bonaccurso},\ and\ \citenamefont
  {Venzmer}}]{Wang2013}%
  \BibitemOpen
  \bibfield  {author} {\bibinfo {author} {\bibfnamefont {X.}~\bibnamefont
  {Wang}}, \bibinfo {author} {\bibfnamefont {L.}~\bibnamefont {Chen}}, \bibinfo
  {author} {\bibfnamefont {E.}~\bibnamefont {Bonaccurso}}, \ and\ \bibinfo
  {author} {\bibfnamefont {J.}~\bibnamefont {Venzmer}},\ }\bibfield  {title}
  {\enquote {\bibinfo {title} {Dynamic wetting of hydrophobic polymers by
  aqueous surfactant and superspreader solutions},}\ }\href {\doibase
  10.1021/la403994y} {\bibfield  {journal} {\bibinfo  {journal} {Langmuir}\
  }\textbf {\bibinfo {volume} {29}},\ \bibinfo {pages} {14855--14864} (\bibinfo
  {year} {2013})}\BibitemShut {NoStop}%
\bibitem [{\citenamefont {Scheller}\ and\ \citenamefont
  {Bousfield}(1995)}]{Scheller1995}%
  \BibitemOpen
  \bibfield  {author} {\bibinfo {author} {\bibfnamefont {B.~L.}\ \bibnamefont
  {Scheller}}\ and\ \bibinfo {author} {\bibfnamefont {D.~W.}\ \bibnamefont
  {Bousfield}},\ }\bibfield  {title} {\enquote {\bibinfo {title} {Newtonian
  drop impact with a solid surface},}\ }\href {\doibase 10.1002/aic.690410602}
  {\bibfield  {journal} {\bibinfo  {journal} {AIChE Journal}\ }\textbf
  {\bibinfo {volume} {41}},\ \bibinfo {pages} {1357--1367} (\bibinfo {year}
  {1995})}\BibitemShut {NoStop}%
\bibitem [{\citenamefont {Tang}\ \emph {et~al.}(2017)\citenamefont {Tang},
  \citenamefont {Qin}, \citenamefont {Weng}, \citenamefont {Zhang},
  \citenamefont {Zhang}, \citenamefont {Li},\ and\ \citenamefont
  {Huang}}]{Tang2017}%
  \BibitemOpen
  \bibfield  {author} {\bibinfo {author} {\bibfnamefont {C.}~\bibnamefont
  {Tang}}, \bibinfo {author} {\bibfnamefont {M.}~\bibnamefont {Qin}}, \bibinfo
  {author} {\bibfnamefont {X.}~\bibnamefont {Weng}}, \bibinfo {author}
  {\bibfnamefont {X.}~\bibnamefont {Zhang}}, \bibinfo {author} {\bibfnamefont
  {P.}~\bibnamefont {Zhang}}, \bibinfo {author} {\bibfnamefont
  {J.}~\bibnamefont {Li}}, \ and\ \bibinfo {author} {\bibfnamefont
  {Z.}~\bibnamefont {Huang}},\ }\bibfield  {title} {\enquote {\bibinfo {title}
  {Dynamics of droplet impact on solid surface with different roughness},}\
  }\href {\doibase 10.1016/j.ijmultiphaseflow.2017.07.002} {\bibfield
  {journal} {\bibinfo  {journal} {International Journal of Multiphase Flow}\
  }\textbf {\bibinfo {volume} {96}},\ \bibinfo {pages} {56 -- 69} (\bibinfo
  {year} {2017})}\BibitemShut {NoStop}%
\bibitem [{\citenamefont {Sen}, \citenamefont {Vaikuntanathan},\ and\
  \citenamefont {Sivakumar}(2014)}]{Sen2014}%
  \BibitemOpen
  \bibfield  {author} {\bibinfo {author} {\bibfnamefont {S.}~\bibnamefont
  {Sen}}, \bibinfo {author} {\bibfnamefont {V.}~\bibnamefont {Vaikuntanathan}},
  \ and\ \bibinfo {author} {\bibfnamefont {D.}~\bibnamefont {Sivakumar}},\
  }\bibfield  {title} {\enquote {\bibinfo {title} {Experimental investigation
  of biofuel drop impact on stainless steel surface},}\ }\href {\doibase
  10.1016/j.expthermflusci.2014.01.014} {\bibfield  {journal} {\bibinfo
  {journal} {Experimental Thermal and Fluid Science}\ }\textbf {\bibinfo
  {volume} {54}},\ \bibinfo {pages} {38--46} (\bibinfo {year}
  {2014})}\BibitemShut {NoStop}%
\bibitem [{\citenamefont {Roisman}(2009)}]{Roisman2009}%
  \BibitemOpen
  \bibfield  {author} {\bibinfo {author} {\bibfnamefont {I.~V.}\ \bibnamefont
  {Roisman}},\ }\bibfield  {title} {\enquote {\bibinfo {title} {Inertia
  dominated drop collisions. ii. an analytical solution of the navier–stokes
  equations for a spreading viscous film},}\ }\href {\doibase
  10.1063/1.3129283} {\bibfield  {journal} {\bibinfo  {journal} {Physics of
  Fluids}\ }\textbf {\bibinfo {volume} {21}},\ \bibinfo {pages} {052104}
  (\bibinfo {year} {2009})}\BibitemShut {NoStop}%
\bibitem [{\citenamefont {Chandra}\ and\ \citenamefont
  {Avedisian}(1991)}]{Chandra1991}%
  \BibitemOpen
  \bibfield  {author} {\bibinfo {author} {\bibfnamefont {S.}~\bibnamefont
  {Chandra}}\ and\ \bibinfo {author} {\bibfnamefont {C.~T.}\ \bibnamefont
  {Avedisian}},\ }\bibfield  {title} {\enquote {\bibinfo {title} {On the
  collision of a droplet with a solid surface},}\ }\href {\doibase
  10.1098/rspa.1991.0002} {\bibfield  {journal} {\bibinfo  {journal}
  {Proceedings of the Royal Society of London. Series A: Mathematical and
  Physical Sciences}\ }\textbf {\bibinfo {volume} {432}},\ \bibinfo {pages}
  {13--41} (\bibinfo {year} {1991})}\BibitemShut {NoStop}%
\bibitem [{\citenamefont {Pasandideh-Fard}\ \emph {et~al.}(1996)\citenamefont
  {Pasandideh-Fard}, \citenamefont {Qiao}, \citenamefont {Chandra},\ and\
  \citenamefont {Mostaghimi}}]{Pasandideh1996}%
  \BibitemOpen
  \bibfield  {author} {\bibinfo {author} {\bibfnamefont {M.}~\bibnamefont
  {Pasandideh-Fard}}, \bibinfo {author} {\bibfnamefont {Y.~M.}\ \bibnamefont
  {Qiao}}, \bibinfo {author} {\bibfnamefont {S.}~\bibnamefont {Chandra}}, \
  and\ \bibinfo {author} {\bibfnamefont {J.}~\bibnamefont {Mostaghimi}},\
  }\bibfield  {title} {\enquote {\bibinfo {title} {Capillary effects during
  droplet impact on a solid surface},}\ }\href {\doibase 10.1063/1.868850}
  {\bibfield  {journal} {\bibinfo  {journal} {Physics of Fluids}\ }\textbf
  {\bibinfo {volume} {8}},\ \bibinfo {pages} {650--659} (\bibinfo {year}
  {1996})}\BibitemShut {NoStop}%
\bibitem [{\citenamefont {Ukiwe}\ and\ \citenamefont {Kwok}(2005)}]{Ukiwe2005}%
  \BibitemOpen
  \bibfield  {author} {\bibinfo {author} {\bibfnamefont {C.}~\bibnamefont
  {Ukiwe}}\ and\ \bibinfo {author} {\bibfnamefont {D.~Y.}\ \bibnamefont
  {Kwok}},\ }\bibfield  {title} {\enquote {\bibinfo {title} {On the maximum
  spreading diameter of impacting droplets on well-prepared solid surfaces},}\
  }\href {\doibase 10.1021/la0481288} {\bibfield  {journal} {\bibinfo
  {journal} {Langmuir}\ }\textbf {\bibinfo {volume} {21}},\ \bibinfo {pages}
  {666--673} (\bibinfo {year} {2005})}\BibitemShut {NoStop}%
\bibitem [{\citenamefont {Du}\ \emph {et~al.}(2021{\natexlab{a}})\citenamefont
  {Du}, \citenamefont {Wang}, \citenamefont {Li}, \citenamefont {Min},\ and\
  \citenamefont {Wu}}]{Du2021_langmuir}%
  \BibitemOpen
  \bibfield  {author} {\bibinfo {author} {\bibfnamefont {J.}~\bibnamefont
  {Du}}, \bibinfo {author} {\bibfnamefont {X.}~\bibnamefont {Wang}}, \bibinfo
  {author} {\bibfnamefont {Y.}~\bibnamefont {Li}}, \bibinfo {author}
  {\bibfnamefont {Q.}~\bibnamefont {Min}}, \ and\ \bibinfo {author}
  {\bibfnamefont {X.}~\bibnamefont {Wu}},\ }\bibfield  {title} {\enquote
  {\bibinfo {title} {Analytical consideration for the maximum spreading factor
  of liquid droplet impact on a smooth solid surface},}\ }\href {\doibase
  10.1021/acs.langmuir.1c01076} {\bibfield  {journal} {\bibinfo  {journal}
  {Langmuir}\ }\textbf {\bibinfo {volume} {37}},\ \bibinfo {pages} {7582--7590}
  (\bibinfo {year} {2021}{\natexlab{a}})}\BibitemShut {NoStop}%
\bibitem [{\citenamefont {Li}, \citenamefont {Ashgriz},\ and\ \citenamefont
  {Chandra}(2010)}]{Li2010}%
  \BibitemOpen
  \bibfield  {author} {\bibinfo {author} {\bibfnamefont {R.}~\bibnamefont
  {Li}}, \bibinfo {author} {\bibfnamefont {N.}~\bibnamefont {Ashgriz}}, \ and\
  \bibinfo {author} {\bibfnamefont {S.}~\bibnamefont {Chandra}},\ }\bibfield
  {title} {\enquote {\bibinfo {title} {{Maximum Spread of Droplet on Solid
  Surface: Low Reynolds and Weber Numbers}},}\ }\href {\doibase
  10.1115/1.4001695} {\bibfield  {journal} {\bibinfo  {journal} {Journal of
  Fluids Engineering}\ }\textbf {\bibinfo {volume} {132}} (\bibinfo {year}
  {2010}),\ 10.1115/1.4001695}\BibitemShut {NoStop}%
\bibitem [{\citenamefont {Park}\ \emph {et~al.}(2003)\citenamefont {Park},
  \citenamefont {Carr}, \citenamefont {Zhu},\ and\ \citenamefont
  {Morris}}]{Park2003}%
  \BibitemOpen
  \bibfield  {author} {\bibinfo {author} {\bibfnamefont {H.}~\bibnamefont
  {Park}}, \bibinfo {author} {\bibfnamefont {W.~W.}\ \bibnamefont {Carr}},
  \bibinfo {author} {\bibfnamefont {J.}~\bibnamefont {Zhu}}, \ and\ \bibinfo
  {author} {\bibfnamefont {J.~F.}\ \bibnamefont {Morris}},\ }\bibfield  {title}
  {\enquote {\bibinfo {title} {Single drop impaction on a solid surface},}\
  }\href {\doibase 10.1002/aic.690491003} {\bibfield  {journal} {\bibinfo
  {journal} {AIChE Journal}\ }\textbf {\bibinfo {volume} {49}},\ \bibinfo
  {pages} {2461--2471} (\bibinfo {year} {2003})}\BibitemShut {NoStop}%
\bibitem [{\citenamefont {Roisman}, \citenamefont {Rioboo},\ and\ \citenamefont
  {Tropea}(2002)}]{Roisman2002}%
  \BibitemOpen
  \bibfield  {author} {\bibinfo {author} {\bibfnamefont {I.~V.}\ \bibnamefont
  {Roisman}}, \bibinfo {author} {\bibfnamefont {R.}~\bibnamefont {Rioboo}}, \
  and\ \bibinfo {author} {\bibfnamefont {C.}~\bibnamefont {Tropea}},\
  }\bibfield  {title} {\enquote {\bibinfo {title} {Normal impact of a liquid
  drop on a dry surface: model for spreading and receding},}\ }\href {\doibase
  10.1098/rspa.2001.0923} {\bibfield  {journal} {\bibinfo  {journal}
  {Proceedings of the Royal Society of London. Series A: Mathematical, Physical
  and Engineering Sciences}\ }\textbf {\bibinfo {volume} {458}},\ \bibinfo
  {pages} {1411--1430} (\bibinfo {year} {2002})}\BibitemShut {NoStop}%
\bibitem [{\citenamefont {Srivastava}\ and\ \citenamefont
  {Kondaraju}(2020)}]{Srivastava2020}%
  \BibitemOpen
  \bibfield  {author} {\bibinfo {author} {\bibfnamefont {T.}~\bibnamefont
  {Srivastava}}\ and\ \bibinfo {author} {\bibfnamefont {S.}~\bibnamefont
  {Kondaraju}},\ }\bibfield  {title} {\enquote {\bibinfo {title} {Analytical
  model for predicting maximum spread of droplet impinging on solid
  surfaces},}\ }\href {\doibase 10.1063/5.0020219} {\bibfield  {journal}
  {\bibinfo  {journal} {Physics of Fluids}\ }\textbf {\bibinfo {volume} {32}},\
  \bibinfo {pages} {092103} (\bibinfo {year} {2020})}\BibitemShut {NoStop}%
\bibitem [{\citenamefont {Wang}\ \emph {et~al.}(2019)\citenamefont {Wang},
  \citenamefont {Yang}, \citenamefont {Wang}, \citenamefont {Zhu},\ and\
  \citenamefont {Fang}}]{Wang2019}%
  \BibitemOpen
  \bibfield  {author} {\bibinfo {author} {\bibfnamefont {F.}~\bibnamefont
  {Wang}}, \bibinfo {author} {\bibfnamefont {L.}~\bibnamefont {Yang}}, \bibinfo
  {author} {\bibfnamefont {L.}~\bibnamefont {Wang}}, \bibinfo {author}
  {\bibfnamefont {Y.}~\bibnamefont {Zhu}}, \ and\ \bibinfo {author}
  {\bibfnamefont {T.}~\bibnamefont {Fang}},\ }\bibfield  {title} {\enquote
  {\bibinfo {title} {Maximum spread of droplet impacting onto solid surfaces
  with different wettabilities: Adopting a rim--lamella shape},}\ }\href
  {\doibase 10.1021/acs.langmuir.8b03748} {\bibfield  {journal} {\bibinfo
  {journal} {Langmuir}\ }\textbf {\bibinfo {volume} {35}},\ \bibinfo {pages}
  {3204--3214} (\bibinfo {year} {2019})}\BibitemShut {NoStop}%
\bibitem [{\citenamefont {Du}\ \emph {et~al.}(2021{\natexlab{b}})\citenamefont
  {Du}, \citenamefont {Chamakos}, \citenamefont {Papathanasiou},\ and\
  \citenamefont {Min}}]{Du2021}%
  \BibitemOpen
  \bibfield  {author} {\bibinfo {author} {\bibfnamefont {J.}~\bibnamefont
  {Du}}, \bibinfo {author} {\bibfnamefont {N.~T.}\ \bibnamefont {Chamakos}},
  \bibinfo {author} {\bibfnamefont {A.~G.}\ \bibnamefont {Papathanasiou}}, \
  and\ \bibinfo {author} {\bibfnamefont {Q.}~\bibnamefont {Min}},\ }\bibfield
  {title} {\enquote {\bibinfo {title} {Initial spreading dynamics of a liquid
  droplet: The effects of wettability, liquid properties, and substrate
  topography},}\ }\href {\doibase 10.1063/5.0049409} {\bibfield  {journal}
  {\bibinfo  {journal} {Physics of Fluids}\ }\textbf {\bibinfo {volume} {33}},\
  \bibinfo {pages} {042118} (\bibinfo {year} {2021}{\natexlab{b}})}\BibitemShut
  {NoStop}%
\bibitem [{\citenamefont {Chamakos}\ \emph {et~al.}(2016)\citenamefont
  {Chamakos}, \citenamefont {Kavousanakis}, \citenamefont {Boudouvis},\ and\
  \citenamefont {Papathanasiou}}]{Chamakos2016}%
  \BibitemOpen
  \bibfield  {author} {\bibinfo {author} {\bibfnamefont {N.~T.}\ \bibnamefont
  {Chamakos}}, \bibinfo {author} {\bibfnamefont {M.~E.}\ \bibnamefont
  {Kavousanakis}}, \bibinfo {author} {\bibfnamefont {A.~G.}\ \bibnamefont
  {Boudouvis}}, \ and\ \bibinfo {author} {\bibfnamefont {A.~G.}\ \bibnamefont
  {Papathanasiou}},\ }\bibfield  {title} {\enquote {\bibinfo {title} {Droplet
  spreading on rough surfaces: Tackling the contact line boundary condition},}\
  }\href {\doibase 10.1063/1.4941577} {\bibfield  {journal} {\bibinfo
  {journal} {Physics of Fluids}\ }\textbf {\bibinfo {volume} {28}},\ \bibinfo
  {pages} {022105} (\bibinfo {year} {2016})}\BibitemShut {NoStop}%
\bibitem [{\citenamefont {Yonemoto}\ and\ \citenamefont
  {Kunugi}(2017)}]{Yonemoto2017}%
  \BibitemOpen
  \bibfield  {author} {\bibinfo {author} {\bibfnamefont {Y.}~\bibnamefont
  {Yonemoto}}\ and\ \bibinfo {author} {\bibfnamefont {T.}~\bibnamefont
  {Kunugi}},\ }\bibfield  {title} {\enquote {\bibinfo {title} {Analytical
  consideration of liquid droplet impingement on solid surfaces},}\ }\href
  {\doibase 10.1038/s41598-017-02450-4} {\bibfield  {journal} {\bibinfo
  {journal} {Scientific Reports}\ }\textbf {\bibinfo {volume} {7}},\ \bibinfo
  {pages} {2362} (\bibinfo {year} {2017})}\BibitemShut {NoStop}%
\bibitem [{\citenamefont {Zhao}\ \emph {et~al.}(2017)\citenamefont {Zhao},
  \citenamefont {Wang}, \citenamefont {Zhang}, \citenamefont {Chen},\ and\
  \citenamefont {Deng}}]{Zhao2017}%
  \BibitemOpen
  \bibfield  {author} {\bibinfo {author} {\bibfnamefont {B.}~\bibnamefont
  {Zhao}}, \bibinfo {author} {\bibfnamefont {X.}~\bibnamefont {Wang}}, \bibinfo
  {author} {\bibfnamefont {K.}~\bibnamefont {Zhang}}, \bibinfo {author}
  {\bibfnamefont {L.}~\bibnamefont {Chen}}, \ and\ \bibinfo {author}
  {\bibfnamefont {X.}~\bibnamefont {Deng}},\ }\bibfield  {title} {\enquote
  {\bibinfo {title} {Impact of viscous droplets on superamphiphobic
  surfaces},}\ }\href {\doibase 10.1021/acs.langmuir.6b03862} {\bibfield
  {journal} {\bibinfo  {journal} {Langmuir}\ }\textbf {\bibinfo {volume}
  {33}},\ \bibinfo {pages} {144--151} (\bibinfo {year} {2017})}\BibitemShut
  {NoStop}%
\bibitem [{\citenamefont {Gao}\ and\ \citenamefont {Li}(2014)}]{Gao2014}%
  \BibitemOpen
  \bibfield  {author} {\bibinfo {author} {\bibfnamefont {X.}~\bibnamefont
  {Gao}}\ and\ \bibinfo {author} {\bibfnamefont {R.}~\bibnamefont {Li}},\
  }\bibfield  {title} {\enquote {\bibinfo {title} {Spread and recoiling of
  liquid droplets impacting solid surfaces},}\ }\href {\doibase
  10.1002/aic.14440} {\bibfield  {journal} {\bibinfo  {journal} {AIChE
  Journal}\ }\textbf {\bibinfo {volume} {60}},\ \bibinfo {pages} {2683--2691}
  (\bibinfo {year} {2014})}\BibitemShut {NoStop}%
\bibitem [{\citenamefont {Mao}, \citenamefont {Kuhn},\ and\ \citenamefont
  {Tran}(1997)}]{Mao1997}%
  \BibitemOpen
  \bibfield  {author} {\bibinfo {author} {\bibfnamefont {T.}~\bibnamefont
  {Mao}}, \bibinfo {author} {\bibfnamefont {D.~C.~S.}\ \bibnamefont {Kuhn}}, \
  and\ \bibinfo {author} {\bibfnamefont {H.}~\bibnamefont {Tran}},\ }\bibfield
  {title} {\enquote {\bibinfo {title} {Spread and rebound of liquid droplets
  upon impact on flat surfaces},}\ }\href {\doibase 10.1002/aic.690430903}
  {\bibfield  {journal} {\bibinfo  {journal} {AIChE Journal}\ }\textbf
  {\bibinfo {volume} {43}},\ \bibinfo {pages} {2169--2179} (\bibinfo {year}
  {1997})}\BibitemShut {NoStop}%
\bibitem [{\citenamefont {Lee}\ \emph {et~al.}(2016)\citenamefont {Lee},
  \citenamefont {Derome}, \citenamefont {Guyer},\ and\ \citenamefont
  {Carmeliet}}]{Lee2016_langmuir}%
  \BibitemOpen
  \bibfield  {author} {\bibinfo {author} {\bibfnamefont {J.~B.}\ \bibnamefont
  {Lee}}, \bibinfo {author} {\bibfnamefont {D.}~\bibnamefont {Derome}},
  \bibinfo {author} {\bibfnamefont {R.}~\bibnamefont {Guyer}}, \ and\ \bibinfo
  {author} {\bibfnamefont {J.}~\bibnamefont {Carmeliet}},\ }\bibfield  {title}
  {\enquote {\bibinfo {title} {Modeling the maximum spreading of liquid
  droplets impacting wetting and nonwetting surfaces},}\ }\href {\doibase
  10.1021/acs.langmuir.5b04557} {\bibfield  {journal} {\bibinfo  {journal}
  {Langmuir}\ }\textbf {\bibinfo {volume} {32}},\ \bibinfo {pages} {1299--1308}
  (\bibinfo {year} {2016})}\BibitemShut {NoStop}%
\bibitem [{\citenamefont {Huang}\ and\ \citenamefont {Chen}(2018)}]{Huang2018}%
  \BibitemOpen
  \bibfield  {author} {\bibinfo {author} {\bibfnamefont {H.-M.}\ \bibnamefont
  {Huang}}\ and\ \bibinfo {author} {\bibfnamefont {X.-P.}\ \bibnamefont
  {Chen}},\ }\bibfield  {title} {\enquote {\bibinfo {title} {Energetic analysis
  of drop’s maximum spreading on solid surface with low impact speed},}\
  }\href {\doibase 10.1063/1.5006439} {\bibfield  {journal} {\bibinfo
  {journal} {Physics of Fluids}\ }\textbf {\bibinfo {volume} {30}},\ \bibinfo
  {pages} {022106} (\bibinfo {year} {2018})}\BibitemShut {NoStop}%
\bibitem [{\citenamefont {Gao}\ \emph {et~al.}(2018)\citenamefont {Gao},
  \citenamefont {Liao}, \citenamefont {Liu},\ and\ \citenamefont
  {Liu}}]{Gao2018}%
  \BibitemOpen
  \bibfield  {author} {\bibinfo {author} {\bibfnamefont {S.}~\bibnamefont
  {Gao}}, \bibinfo {author} {\bibfnamefont {Q.}~\bibnamefont {Liao}}, \bibinfo
  {author} {\bibfnamefont {W.}~\bibnamefont {Liu}}, \ and\ \bibinfo {author}
  {\bibfnamefont {Z.}~\bibnamefont {Liu}},\ }\bibfield  {title} {\enquote
  {\bibinfo {title} {Nanodroplets impact on rough surfaces: A simulation and
  theoretical study},}\ }\href {\doibase 10.1021/acs.langmuir.8b00480}
  {\bibfield  {journal} {\bibinfo  {journal} {Langmuir}\ }\textbf {\bibinfo
  {volume} {34}},\ \bibinfo {pages} {5910--5917} (\bibinfo {year}
  {2018})}\BibitemShut {NoStop}%
\bibitem [{\citenamefont {Qin}\ \emph {et~al.}(2019)\citenamefont {Qin},
  \citenamefont {Tang}, \citenamefont {Tong}, \citenamefont {Zhang},\ and\
  \citenamefont {Huang}}]{Qin2019}%
  \BibitemOpen
  \bibfield  {author} {\bibinfo {author} {\bibfnamefont {M.}~\bibnamefont
  {Qin}}, \bibinfo {author} {\bibfnamefont {C.}~\bibnamefont {Tang}}, \bibinfo
  {author} {\bibfnamefont {S.}~\bibnamefont {Tong}}, \bibinfo {author}
  {\bibfnamefont {P.}~\bibnamefont {Zhang}}, \ and\ \bibinfo {author}
  {\bibfnamefont {Z.}~\bibnamefont {Huang}},\ }\bibfield  {title} {\enquote
  {\bibinfo {title} {On the role of liquid viscosity in affecting droplet
  spreading on a smooth solid surface},}\ }\href {\doibase
  10.1016/j.ijmultiphaseflow.2019.05.002} {\bibfield  {journal} {\bibinfo
  {journal} {International Journal of Multiphase Flow}\ }\textbf {\bibinfo
  {volume} {117}},\ \bibinfo {pages} {53--63} (\bibinfo {year}
  {2019})}\BibitemShut {NoStop}%
\bibitem [{\citenamefont {Clanet}\ \emph {et~al.}(2004)\citenamefont {Clanet},
  \citenamefont {Beguin}, \citenamefont {Richard},\ and\ \citenamefont
  {Quere}}]{Clanet2004}%
  \BibitemOpen
  \bibfield  {author} {\bibinfo {author} {\bibfnamefont {C.}~\bibnamefont
  {Clanet}}, \bibinfo {author} {\bibfnamefont {C.}~\bibnamefont {Beguin}},
  \bibinfo {author} {\bibfnamefont {D.}~\bibnamefont {Richard}}, \ and\
  \bibinfo {author} {\bibfnamefont {D.}~\bibnamefont {Quere}},\ }\bibfield
  {title} {\enquote {\bibinfo {title} {Maximal deformation of an impacting
  drop},}\ }\href {\doibase 10.1017/S0022112004000904} {\bibfield  {journal}
  {\bibinfo  {journal} {Journal of Fluid Mechanics}\ }\textbf {\bibinfo
  {volume} {517}},\ \bibinfo {pages} {199--208} (\bibinfo {year}
  {2004})}\BibitemShut {NoStop}%
\bibitem [{\citenamefont {Bayer}\ and\ \citenamefont
  {Megaridis}(2006)}]{Bayer2006}%
  \BibitemOpen
  \bibfield  {author} {\bibinfo {author} {\bibfnamefont {I.~S.}\ \bibnamefont
  {Bayer}}\ and\ \bibinfo {author} {\bibfnamefont {C.~M.}\ \bibnamefont
  {Megaridis}},\ }\bibfield  {title} {\enquote {\bibinfo {title} {Contact angle
  dynamics in droplets impacting on flat surfaces with different wetting
  characteristics},}\ }\href {\doibase 10.1017/S0022112006000231} {\bibfield
  {journal} {\bibinfo  {journal} {Journal of Fluid Mechanics}\ }\textbf
  {\bibinfo {volume} {558}},\ \bibinfo {pages} {415–449} (\bibinfo {year}
  {2006})}\BibitemShut {NoStop}%
\bibitem [{\citenamefont {Seo}\ \emph {et~al.}(2015)\citenamefont {Seo},
  \citenamefont {Lee}, \citenamefont {Kim},\ and\ \citenamefont
  {Yoon}}]{Seo2015}%
  \BibitemOpen
  \bibfield  {author} {\bibinfo {author} {\bibfnamefont {J.}~\bibnamefont
  {Seo}}, \bibinfo {author} {\bibfnamefont {J.~S.}\ \bibnamefont {Lee}},
  \bibinfo {author} {\bibfnamefont {H.~Y.}\ \bibnamefont {Kim}}, \ and\
  \bibinfo {author} {\bibfnamefont {S.~S.}\ \bibnamefont {Yoon}},\ }\bibfield
  {title} {\enquote {\bibinfo {title} {Empirical model for the maximum
  spreading diameter of low-viscosity droplets on a dry wall},}\ }\href
  {\doibase 10.1016/j.expthermflusci.2014.10.019} {\bibfield  {journal}
  {\bibinfo  {journal} {Experimental Thermal and Fluid Science}\ }\textbf
  {\bibinfo {volume} {61}},\ \bibinfo {pages} {121--129} (\bibinfo {year}
  {2015})}\BibitemShut {NoStop}%
\bibitem [{\citenamefont {Aksoy}\ \emph {et~al.}(2022)\citenamefont {Aksoy},
  \citenamefont {Eneren}, \citenamefont {Koos},\ and\ \citenamefont
  {Vetrano}}]{Aksoy2022}%
  \BibitemOpen
  \bibfield  {author} {\bibinfo {author} {\bibfnamefont {Y.}~\bibnamefont
  {Aksoy}}, \bibinfo {author} {\bibfnamefont {P.}~\bibnamefont {Eneren}},
  \bibinfo {author} {\bibfnamefont {E.}~\bibnamefont {Koos}}, \ and\ \bibinfo
  {author} {\bibfnamefont {M.}~\bibnamefont {Vetrano}},\ }\bibfield  {title}
  {\enquote {\bibinfo {title} {Spreading-splashing transition of nanofluid
  droplets on a smooth flat surface},}\ }\href {\doibase
  10.1016/j.jcis.2021.07.157} {\bibfield  {journal} {\bibinfo  {journal}
  {Journal of Colloid and Interface Science}\ }\textbf {\bibinfo {volume}
  {606}},\ \bibinfo {pages} {434--443} (\bibinfo {year} {2022})}\BibitemShut
  {NoStop}%
\bibitem [{\citenamefont {Palacios}\ \emph {et~al.}(2013)\citenamefont
  {Palacios}, \citenamefont {Hernandez}, \citenamefont {Gomez}, \citenamefont
  {Zanzi},\ and\ \citenamefont {Lopez}}]{Palacios2013}%
  \BibitemOpen
  \bibfield  {author} {\bibinfo {author} {\bibfnamefont {J.}~\bibnamefont
  {Palacios}}, \bibinfo {author} {\bibfnamefont {J.}~\bibnamefont {Hernandez}},
  \bibinfo {author} {\bibfnamefont {P.}~\bibnamefont {Gomez}}, \bibinfo
  {author} {\bibfnamefont {C.}~\bibnamefont {Zanzi}}, \ and\ \bibinfo {author}
  {\bibfnamefont {J.}~\bibnamefont {Lopez}},\ }\bibfield  {title} {\enquote
  {\bibinfo {title} {Experimental study of splashing patterns and the
  splashing/deposition threshold in drop impacts onto dry smooth solid
  surfaces},}\ }\href {\doibase 10.1016/j.expthermflusci.2012.08.020}
  {\bibfield  {journal} {\bibinfo  {journal} {Experimental Thermal and Fluid
  Science}\ }\textbf {\bibinfo {volume} {44}},\ \bibinfo {pages} {571 -- 582}
  (\bibinfo {year} {2013})}\BibitemShut {NoStop}%
\bibitem [{\citenamefont {Lin}\ \emph {et~al.}(2018)\citenamefont {Lin},
  \citenamefont {Zhao}, \citenamefont {Zou}, \citenamefont {Guo}, \citenamefont
  {Wei},\ and\ \citenamefont {Chen}}]{Lin2018}%
  \BibitemOpen
  \bibfield  {author} {\bibinfo {author} {\bibfnamefont {S.}~\bibnamefont
  {Lin}}, \bibinfo {author} {\bibfnamefont {B.}~\bibnamefont {Zhao}}, \bibinfo
  {author} {\bibfnamefont {S.}~\bibnamefont {Zou}}, \bibinfo {author}
  {\bibfnamefont {J.}~\bibnamefont {Guo}}, \bibinfo {author} {\bibfnamefont
  {Z.}~\bibnamefont {Wei}}, \ and\ \bibinfo {author} {\bibfnamefont
  {L.}~\bibnamefont {Chen}},\ }\bibfield  {title} {\enquote {\bibinfo {title}
  {Impact of viscous droplets on different wettable surfaces: Impact phenomena,
  the maximum spreading factor, spreading time and post-impact oscillation},}\
  }\href {\doibase 10.1016/j.jcis.2017.12.086} {\bibfield  {journal} {\bibinfo
  {journal} {Journal of Colloid and Interface Science}\ }\textbf {\bibinfo
  {volume} {516}},\ \bibinfo {pages} {86--97} (\bibinfo {year}
  {2018})}\BibitemShut {NoStop}%
\bibitem [{\citenamefont {Wang}, \citenamefont {Chen},\ and\ \citenamefont
  {Bonaccurso}(2015)}]{Wang2015}%
  \BibitemOpen
  \bibfield  {author} {\bibinfo {author} {\bibfnamefont {X.}~\bibnamefont
  {Wang}}, \bibinfo {author} {\bibfnamefont {L.}~\bibnamefont {Chen}}, \ and\
  \bibinfo {author} {\bibfnamefont {E.}~\bibnamefont {Bonaccurso}},\ }\bibfield
   {title} {\enquote {\bibinfo {title} {Comparison of spontaneous wetting and
  drop impact dynamics of aqueous surfactant solutions on hydrophobic
  polypropylene surfaces: scaling of the contact radius},}\ }\href {\doibase
  10.1007/s00396-014-3410-x} {\bibfield  {journal} {\bibinfo  {journal}
  {Colloid and Polymer Science}\ }\textbf {\bibinfo {volume} {293}},\ \bibinfo
  {pages} {257--265} (\bibinfo {year} {2015})}\BibitemShut {NoStop}%
\bibitem [{\citenamefont {de~Ruiter}, \citenamefont {Pepper},\ and\
  \citenamefont {Stone}(2010)}]{Ruiter2010}%
  \BibitemOpen
  \bibfield  {author} {\bibinfo {author} {\bibfnamefont {J.}~\bibnamefont
  {de~Ruiter}}, \bibinfo {author} {\bibfnamefont {R.~E.}\ \bibnamefont
  {Pepper}}, \ and\ \bibinfo {author} {\bibfnamefont {H.~A.}\ \bibnamefont
  {Stone}},\ }\bibfield  {title} {\enquote {\bibinfo {title} {Thickness of the
  rim of an expanding lamella near the splash threshold},}\ }\href {\doibase
  10.1063/1.3313360} {\bibfield  {journal} {\bibinfo  {journal} {Physics of
  Fluids}\ }\textbf {\bibinfo {volume} {22}},\ \bibinfo {pages} {022104}
  (\bibinfo {year} {2010})}\BibitemShut {NoStop}%
\bibitem [{\citenamefont {Pierce}, \citenamefont {Carmona},\ and\ \citenamefont
  {Amirfazli}(2008)}]{Pierce2008}%
  \BibitemOpen
  \bibfield  {author} {\bibinfo {author} {\bibfnamefont {E.}~\bibnamefont
  {Pierce}}, \bibinfo {author} {\bibfnamefont {F.}~\bibnamefont {Carmona}}, \
  and\ \bibinfo {author} {\bibfnamefont {A.}~\bibnamefont {Amirfazli}},\
  }\bibfield  {title} {\enquote {\bibinfo {title} {Understanding of sliding and
  contact angle results in tilted plate experiments},}\ }\href {\doibase
  10.1016/j.colsurfa.2007.09.032} {\bibfield  {journal} {\bibinfo  {journal}
  {Colloids and Surfaces A: Physicochemical and Engineering Aspects}\ }\textbf
  {\bibinfo {volume} {323}},\ \bibinfo {pages} {73--82} (\bibinfo {year}
  {2008})},\ \bibinfo {note} {bubble and Drop Interfaces}\BibitemShut {NoStop}%
\bibitem [{\citenamefont {Almohammadi}\ and\ \citenamefont
  {Amirfazli}(2019)}]{Almohammadi2019}%
  \BibitemOpen
  \bibfield  {author} {\bibinfo {author} {\bibfnamefont {H.}~\bibnamefont
  {Almohammadi}}\ and\ \bibinfo {author} {\bibfnamefont {A.}~\bibnamefont
  {Amirfazli}},\ }\bibfield  {title} {\enquote {\bibinfo {title} {Droplet
  impact: Viscosity and wettability effects on splashing},}\ }\href {\doibase
  10.1016/j.jcis.2019.05.101} {\bibfield  {journal} {\bibinfo  {journal}
  {Journal of Colloid and Interface Science}\ }\textbf {\bibinfo {volume}
  {553}},\ \bibinfo {pages} {22 -- 30} (\bibinfo {year} {2019})}\BibitemShut
  {NoStop}%
\bibitem [{\citenamefont {Jadidbonab}\ \emph {et~al.}(2018)\citenamefont
  {Jadidbonab}, \citenamefont {Malgarinos}, \citenamefont {Karathanassis},
  \citenamefont {Mitroglou},\ and\ \citenamefont {Gavaises}}]{Jadidbonab2018}%
  \BibitemOpen
  \bibfield  {author} {\bibinfo {author} {\bibfnamefont {H.}~\bibnamefont
  {Jadidbonab}}, \bibinfo {author} {\bibfnamefont {I.}~\bibnamefont
  {Malgarinos}}, \bibinfo {author} {\bibfnamefont {I.}~\bibnamefont
  {Karathanassis}}, \bibinfo {author} {\bibfnamefont {N.}~\bibnamefont
  {Mitroglou}}, \ and\ \bibinfo {author} {\bibfnamefont {M.}~\bibnamefont
  {Gavaises}},\ }\bibfield  {title} {\enquote {\bibinfo {title} {<i>we</i>-t
  classification of diesel fuel droplet impact regimes},}\ }\href {\doibase
  10.1098/rspa.2017.0759} {\bibfield  {journal} {\bibinfo  {journal}
  {Proceedings of the Royal Society A: Mathematical, Physical and Engineering
  Sciences}\ }\textbf {\bibinfo {volume} {474}},\ \bibinfo {pages} {20170759}
  (\bibinfo {year} {2018})}\BibitemShut {NoStop}%
\bibitem [{\citenamefont {Lee}, \citenamefont {Derome},\ and\ \citenamefont
  {Carmeliet}(2016)}]{Lee2016_JCIS}%
  \BibitemOpen
  \bibfield  {author} {\bibinfo {author} {\bibfnamefont {J.}~\bibnamefont
  {Lee}}, \bibinfo {author} {\bibfnamefont {D.}~\bibnamefont {Derome}}, \ and\
  \bibinfo {author} {\bibfnamefont {J.}~\bibnamefont {Carmeliet}},\ }\bibfield
  {title} {\enquote {\bibinfo {title} {Drop impact on natural porous stones},}\
  }\href {\doibase 10.1016/j.jcis.2016.02.008} {\bibfield  {journal} {\bibinfo
  {journal} {Journal of Colloid and Interface Science}\ }\textbf {\bibinfo
  {volume} {469}},\ \bibinfo {pages} {147--156} (\bibinfo {year}
  {2016})}\BibitemShut {NoStop}%
\bibitem [{\citenamefont {Quetzeri-Santiago}, \citenamefont
  {Castrej{\'o}n-Pita},\ and\ \citenamefont
  {Castrej{\'o}n-Pita}(2020)}]{Santiago2020}%
  \BibitemOpen
  \bibfield  {author} {\bibinfo {author} {\bibfnamefont {M.~A.}\ \bibnamefont
  {Quetzeri-Santiago}}, \bibinfo {author} {\bibfnamefont {J.~R.}\ \bibnamefont
  {Castrej{\'o}n-Pita}}, \ and\ \bibinfo {author} {\bibfnamefont {A.~A.}\
  \bibnamefont {Castrej{\'o}n-Pita}},\ }\bibfield  {title} {\enquote {\bibinfo
  {title} {On the analysis of the contact angle for impacting droplets using a
  polynomial fitting approach},}\ }\href {\doibase 10.1007/s00348-020-02971-1}
  {\bibfield  {journal} {\bibinfo  {journal} {Experiments in Fluids}\ }\textbf
  {\bibinfo {volume} {61}},\ \bibinfo {pages} {143} (\bibinfo {year}
  {2020})}\BibitemShut {NoStop}%
\bibitem [{\citenamefont {Sikalo}\ \emph {et~al.}(2002)\citenamefont {Sikalo},
  \citenamefont {Marengo}, \citenamefont {Tropea},\ and\ \citenamefont
  {Ganic}}]{Sikalo2002}%
  \BibitemOpen
  \bibfield  {author} {\bibinfo {author} {\bibfnamefont {S.}~\bibnamefont
  {Sikalo}}, \bibinfo {author} {\bibfnamefont {M.}~\bibnamefont {Marengo}},
  \bibinfo {author} {\bibfnamefont {C.}~\bibnamefont {Tropea}}, \ and\ \bibinfo
  {author} {\bibfnamefont {E.}~\bibnamefont {Ganic}},\ }\bibfield  {title}
  {\enquote {\bibinfo {title} {Analysis of impact of droplets on horizontal
  surfaces},}\ }\href {\doibase 10.1016/S0894-1777(01)00109-1} {\bibfield
  {journal} {\bibinfo  {journal} {Experimental Thermal and Fluid Science}\
  }\textbf {\bibinfo {volume} {25}},\ \bibinfo {pages} {503 -- 510} (\bibinfo
  {year} {2002})},\ \bibinfo {note} {international Thermal Science
  Seminar}\BibitemShut {NoStop}%
\end{thebibliography}%

\end{document}